\documentclass[aps,pra,floatfix,twocolumn,amsmath,amssymb,showpacs,10pt]{revtex4-1}

\usepackage{dcolumn}
\usepackage{graphicx}
\usepackage{pst-all,pstricks-add}

\newcommand{\ket}[1]{\lvert #1 \rangle}           
\newcommand{\bra}[1]{\langle #1 \lvert}           
\newcommand{\innerprod}[2]{\langle #1 | #2 \rangle}           
\newcommand{\expv}[1]{\langle #1 \rangle}           
\newcommand{\myvector}[1]{\boldsymbol{\mathrm{#1}}}
\newcommand{\unitvec}[1]{\boldsymbol{\hat{\mathrm{#1}}}}

\newcommand{\rhoi}{\hat{\rho}_\mathrm{i}}

\newcommand{\rhoo}{\hat{\rho}_\mathrm{0}}
\newcommand{\rhooorder}[1]{\hat{\rho}^{(#1)}_0}
\newcommand{\rhoiorder}[1]{\hat{\rho}^{(#1)}_\mathrm{i}}
\newcommand{\rhoforder}[1]{\hat{\rho}^{(#1)}_\mathrm{f}}

\newcommand{\rhof}{\hat{\rho}_\mathrm{f}}

\newcommand{\rhofderiv}{\hat{\dot{\rho}}_\mathrm{f}}
\newcommand{\rhofderivorder}[1]{\hat{\dot{\rho}}^{(#1)}_\mathrm{f}}

\newcommand{\rhom}{\hat{\rho}_\mathrm{m}}
\newcommand{\rhomorder}[1]{\hat{\rho}^{(#1)}_\mathrm{m}}

\newcommand{\rinitial}{\myvector{r}_\textrm{0}}

\newcommand{\rinput}{\myvector{r}_\textrm{i}}
\newcommand{\rfinal}{\myvector{r}_\textrm{f}}

\newcommand{\Horder}[1]{H^{(#1)}}

\newcommand{\Hsopt}{H_\textrm{s opt}}
\newcommand{\Hcorropt}{H_\textrm{corr opt}}

\newcommand{\score}{\hat{L}}
\newcommand{\scoreorder}[1]{\hat{L}^{(#1)}}

\newcommand{\scorestatej}{\ket{\phi^{(j)}}}

\newcommand{\scorestatezero}{\ket{\phi^{(0)}}}
\newcommand{\scorestateone}{\ket{\phi^{(1)}}}

\newcommand{\scoreevalj}{\mu^{(j)}}
\newcommand{\scoreevalkj}{\mu^{(k-j)}}
\newcommand{\scoreevalzero}{\mu^{(0)}}
\newcommand{\scoreevalone}{\mu^{(1)}}

\newcommand{\sigmax}{\hat{\sigma}_x}
\newcommand{\sigmay}{\hat{\sigma}_y}
\newcommand{\sigmaz}{\hat{\sigma}_z}

\newcommand{\sigmagen}[1]{\hat{\sigma}_{\boldsymbol{\mathrm{#1}}}}
\newcommand{\iop}{\hat{I}}
\newcommand{\lambdaest}{\lambda_\mathrm{est}}

\newcommand{\uprep}{\hat{U}_\mathrm{prep}}
\newcommand{\uc}{\hat{U}_{\boldsymbol{\mathrm{c}}}}
\newcommand{\ucconj}{\hat{U}_{\boldsymbol{\mathrm{c}}}^\dagger}
\newcommand{\projc}{P_{\myvector{c}}}
 
\DeclareMathOperator{\Trace}{Tr}
\DeclareMathOperator{\variance}{var}
\DeclareMathOperator{\diag}{diag}

\begin{document}

\author{David Collins}
\affiliation{Department of Physical and Environmental Sciences, Colorado Mesa University, Grand Junction, CO 81501}
\email{dacollin@coloradomesa.edu}


\title{Qubit-channel metrology with very noisy initial states}

\begin{abstract}
 We consider an arbitrary qubit channel depending on a single parameter, which is to be estimated by a physical process. Using the quantum Fisher information per channel invocation to quantify the estimation accuracy, we consider various estimation protocols when the available initial states are mixed with very low purity, $r$. We compare a protocol using a single channel invocation on one out of $n$ qubits prepared in a particular correlated input state to the optimal protocol using uncorrelated input states, with the same initial-state purity. We show that, to lowest order in initial-state purity, for a unital channel this correlated-state protocol enhances the estimation accuracy by a factor between $n-1$ and $n$, provided that $nr^2 \ll 1$. We also show that to lowest order in initial-state purity, a broad class of non-unital channels yields no gain regardless of the input state.    
\end{abstract}


\maketitle


\section{Introduction}
\label{sec:intro}

Quantum parameter estimation, or metrology, considers using physical quantum systems as measuring devices. Typically a system is prepared in a known state and is then subjected to an evolution of a known type but which depends on an unknown parameter that is to be estimated. The parameter must subsequently be inferred from measurements on the system. Classical statistics and quantum physics constrain the success of such procedures; combining these has led to a quantum estimation framework~\cite{helstrom76,caves81,braunstein94,sarovar06,giovannetti06,paris09,kok10,giovannetti11,kolodynski13}. 

This has been applied to various situations, including estimation of parameters in phase-shifts~\cite{giovannetti06}, depolarizing channels~\cite{fujiwara01,sasaki02,frey11} Pauli channels~\cite{fujiwara03,ji08} and amplitude damping channels~\cite{fujiwara04}. A key issue is whether using states only available to quantum systems (such as entangled states) enhances the estimation accuracy compared to that for ``classical'' repeat and average strategies using uncorrelated states. Sometimes this is true.  

Most studies focus on the absolute optimal situations, which require pure initial states. However, in some situations such as solution-state nuclear magnetic resonance (NMR), pure states are unavailable and the issue becomes whether advantages arise when correlating mixed or noisy states that would otherwise be used in an uncorrelated estimation protocol. This has been addressed for the phase-shift~\cite{dariano05,datta11,modi11,pinel13}, phase-flip~\cite{collins13} and depolarizing channels~\cite{collins15}. These studies considered the situation where all available qubits are initially in mixed states and focused on the enhancement of estimation accuracy in terms of the quantum Fisher information. So far there is no general result for all single qubit channels and for all initial state purities; each study yielded a distinct algebraic expression that appears difficult to interpret for a general degree of purity in the initial states. However it was observed that this simplifies considerably for the qubit phase-flip and depolarizing channels~\cite{collins13,collins15} when the initial-state purity is very low and, for a particular protocol which uses correlated states, the accuracy can be enhanced by a factor equal to the number of qubits available. 

The question that this article addresses is whether, for any single qubit channel, there exists a parameter estimation protocol that uses correlated states and provides such accuracy enhancements when the purity of the available initial states is extremely low. While this is partly motivated by the answers from previous studies of specific channels, a physical motivation comes from NMR, where the available purity of the states is very low. A generic model for this could be an ensemble of identical molecules each with the same number of spin-1/2 nuclei and a channel acting on the same nuclear spin in each molecule. Would there be a parameter estimation advantage to correlating the spins within each molecule prior to the channel action?     

We note that the situation where some qubits are initially in pure states and others are in mixed states has been studied and also shows advantages for estimation~\cite{cable16}; we do not consider this case here. 

This article is organized as follows. Section~\ref{sec:qubitchannels} reviews the general description of qubit system evolution and describes the basic notions of parameter estimation. Section~\ref{sec:genestimation} reviews quantum estimation theory and adapts this to situations where the available initial states are very noisy. Section~\ref{sec:sqsc} applies this to a single qubit protocol, which serves as a baseline for comparison with a multiple qubit correlated-state protocol that is described in Sec.~\ref{sec:corrstate}. This contains the main results of this article; these are summarized in Sec.~\ref{sec:summary}.



\section{Single parameter qubit channel metrology}
\label{sec:qubitchannels}

We consider general single qubit quantum channels. Prior to evolution, the \emph{channel input} state for a single qubit can be represented as
\begin{equation}
  \rhoi = \frac{1}{2}\; \left( \iop  + r \sigmagen{\rinput}  \right),
	\label{eq:initialsingle}
\end{equation}
where $\hat{I}$ is the identity operator, $\rinput$ is the \emph{input state Bloch-sphere direction,} a three dimensional real unit vector and $\sigmagen{\rinput} = \rinput \cdot \hat{\myvector{\sigma}} = r_\textrm{i $x$} \sigmax + r_\textrm{i $y$} \sigmay + r_\textrm{i $z$} \sigmaz$ (throughout this article we use the notation $\sigmagen{a} := \myvector{a} \cdot \hat{\myvector{\sigma}}$ where $\myvector{a}$ is any real three dimensional vector). Here $r$, which satisfies $0 \leqslant r \leqslant 1$, is called the purity of the state and quantifies the mixedness or noisiness of the state. Under the channel, $\rhoi \mapsto \rhof$ and, again generically, $\rhof = \left( \iop + \sigmagen{\rfinal} \right)/2$ where $\rfinal$ is the final state Bloch-sphere direction with $\lvert \rfinal \rvert \leqslant 1$. For any channel~\cite{nielsen00},
\begin{equation}
 \rfinal = M(r \rinput) +  \myvector{d} = rM\rinput + \myvector{d}
\label{eq:Blochspherevectorevol}
\end{equation}
where $M$ is $3\times3$ real \emph{Bloch-sphere matrix}, $\myvector{d}$ and a real \emph{Bloch-sphere shift vector}. In Appendix~\ref{app:channelparameterization} we show that $\rvert \myvector{d} \lvert \leqslant 1$ and $\rvert \myvector{d} \lvert=1$ is only possible when $M=0.$

By linearity, the channel maps $\iop \mapsto \iop + \sigmagen{d}$ and also $ \sigmagen{\rinput}  \mapsto M\rinput \cdot \hat{\myvector{\sigma}}.$ 

Qubit channels for which $\myvector{d} = 0$ are called unital; these map $\iop \mapsto \iop.$ Examples are unitary channels, Pauli channels and the depolarizing channel. Non-unital channels, for which $\myvector{d} \neq 0$, include the amplitude damping channel.

We consider channels which depend, via only $M$ and $\myvector{d},$ on a single parameter, $\lambda$, which is independent of the channel input state.  The task will be to estimate the parameter by a physical process in which one or more qubits, prepared in known input states, undergo evolution via one or more identical copies of the channel. The channel actions are followed by measurements, whose outcomes are used to infer the parameter. The goal will be to choose input states, measurements and statistical inference processes that minimize fluctuations in the estimates they generate; we assume that the key cost of such procedures is the number of channel invocations.


\section{Entanglement-assisted metrology with noisy initial states}
\label{sec:genestimation}

A standard formalism for assessing physical quantum estimation uses the density operator, $\rhof(\lambda)$ for the (possibly multiple qubit) entire system immediately after the final channel invocation. The estimate, $\lambdaest$, is inferred from measurement outcomes via a known estimator function. We will require that this estimator is unbiased, i.e.\ the mean of the estimates equals the true parameter value. Then the classical Cram\'{e}r-Rao bound (CRB), bounds the variance in the estimate via $\variance{(\lambdaest)}:= \left< \left( \lambdaest - \expv{\lambdaest} \right)^2 \right> \geqslant 1/F(\lambda)$ regardless of the choice of estimator~\cite{cramer46,paris09} (the angle brackets indicating the mean over all possible measurement outcomes). The classical Fisher information,
\begin{equation}
	F(\lambda):= \int 
	             \left[ 
	               \frac{\partial \ln{p(x_1, x_2 \ldots|\lambda)}}{\partial \lambda}
	             \right]^2\;
	             \;
	             \mathrm{d}x_1 \mathrm{d}x_2 \ldots, 
	\label{eq:classicalfisher}
\end{equation}
is determined from the probability distribution $p(x_1, x_2 \ldots|\lambda)$ for the process measurement outcomes, $x_1, x_2, \ldots$. There is always an estimator which asymptotically attains the lower bound~\cite{cramer46}.

In quantum estimation the choice of measurement affects the probability distribution used to compute the classical Fisher information. However, a further constraint is given by the quantum Cram\'{e}r-Rao bound (QCRB), $F(\lambda) \leqslant H(\lambda)$, where the quantum Fisher information (QFI) is
\begin{equation}
	H(\lambda) = \Trace{\left[ \rhof(\lambda) \score^2(\lambda)\right]},
	\label{eq:quantumfisher}
\end{equation}
and $\score(\lambda)$ is the symmetric logarithmic derivative (SLD) defined via
\begin{equation}
	\frac{\partial \rhof(\lambda)}{\partial \lambda} = \frac{1}{2}\;
	                                                   \left[
	                                                     \score(\lambda)
	                                                     \rhof(\lambda)
	                                                     +
	                                                     \rhof(\lambda)
	                                                     \score(\lambda)
	                                                   \right].
  \label{eq:slddefintion}
\end{equation}
The SLD and the QFI only depend on the pre-measurement system state and thus $\variance{(\lambdaest)} \geqslant 1/H(\lambda)$ offers a bound that is independent of both the choice of measurement and estimator~\cite{braunstein94,paris09,oloan10,nagaoka12ch9,watanabe14}.

The SLD can always be computed from a diagonal decomposition, $\rhof(\lambda) = \sum_j p_j(\lambda) \ket{\phi_j(\lambda)}\bra{\phi_j(\lambda)},$ via~\cite{paris09,kolodynski13}, 
\begin{equation}
  \score(\lambda) = 2 \sum_{j,k} 
	               \frac{\bra{\phi_j} \rhofderiv \ket{\phi_k}}{p_j + p_k}\;
								 \ket{\phi_j}\bra{\phi_j}
	\label{eq:qfieigenval}
\end{equation}
where the dot indicates differentiation with respect to the parameter. In some cases there are simpler algebraic methods for computing the SLD~\cite{collins13}. Also, simple matrix algebra and Eqs.~\eqref{eq:quantumfisher} and~\eqref{eq:slddefintion} give 
\begin{equation}
 H(\lambda) = \Trace{\left[  \frac{\partial \rhof(\lambda)}{\partial \lambda} \score(\lambda) \right]}.
            \label{eq:quantumfishertwo}
\end{equation}

It is always possible to saturate the QCRB by choosing a projective measurement in the eigenbasis of the SLD but it cannot be assured that this choice is independent of the unknown parameter~\cite{paris09,nagaoka12ch9,toscano17}. In such cases, there exist various other measurement schemes that asymptotically saturate the QCRB~\cite{barndorff00}.

Thus the QFI quantifies the accuracy of possible physical measurement procedures. Generally the task in any quantum parameter estimation study has been to engineer a final system state that maximizes the QFI, subject to various system constraints and resources (i.e.\ number of channel invocations, number of available systems, types of initial states available, \ldots)~\cite{braunstein94,fujiwara01,fujiwara03,ballester04,fujiwara04,hotta05,giovannetti06,sarovar06,ji08,paris09,oloan10,datta11,escher11,frey11,giovannetti11,modi11,zwierz12,collins13,kolodynski13,demkowicz14,collins15}. We will adopt the common approach in which the only costly resource is the number of channel invocations. We therefore aim to evaluate estimation protocols via the QFI per channel invocation.

\begin{figure}%
 \includegraphics[scale=.6]{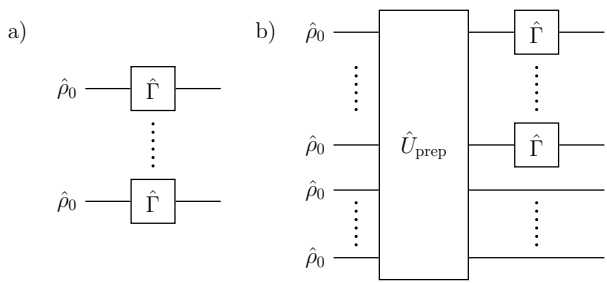}	
\caption{Two quantum metrology protocols. a) A ``classical'' independent channel invocation protocol in which the channel $\hat{\Gamma}(\lambda)$ is invoked on $m$ systems each in the same input state, $\rhoo$. b) An entanglement-assisted protocol using $n$ systems, prepared in an entangled or correlated state via a multi-system preparatory unitary $\uprep$ and with the channel invoked once on each of $m$ of these. The lower $n-m$ systems serve as ancillas.    
				\label{fig:genscheme}%
				}
\end{figure}

A protocol that uses multiple copies of the channel might enhance the QFI. Indeed, an \emph{independent channel invocation protocol} which invokes the channel once on each of $m$ systems prepared independently in the same input state, as illustrated in Fig.~\ref{fig:genscheme}a), gives exactly an $m$-fold increase in the QFI~\cite{oloan10}. Here the QFI per channel invocation is the same as a procedure in which the channel is invoked once on a single system.

In contrast, entanglement-assisted metrology considers protocols, illustrated in Fig~\ref{fig:genscheme}b), where the available quantum systems are prepared in an entangled or otherwise correlated state and thereafter the channel is invoked on a subset of these while the remaining ancillary systems function as spectators in a noiseless environment.  Such entanglement assistance can enhance the QFI per channel invocation (versus uncorrelated or independent protocols)~\cite{kolodynski13,demkowicz14,dur14,huang16,capallero05}.  A key issue in quantum metrology is to establish when and to what extent entanglement assisted protocols can assist parameter estimation. 

\begin{figure}[h]%
 \includegraphics[scale=.6]{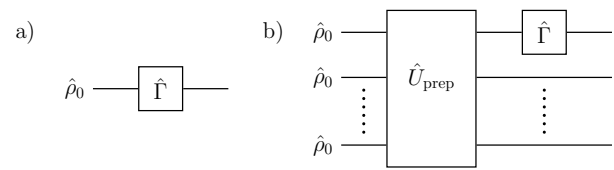}	
\caption{Single channel qubit metrology protocols. a) The SQSC protocol with a single channel invocation on a single qubit. b) An entanglement-assisted protocol using $n$ qubits with the channel invoked once on one of them.    
				\label{fig:singlechannelschemes}%
				}
\end{figure}

We address this for parameter estimation for single qubit channels subject to the following considerations. First, we assume that a fixed finite number of qubits, $n$, is available and each is initially in the same \emph{initial state}; generically this can be expressed as $\rhoo = (\iop + r \sigmagen{\rinitial} )/2$ where $\rinitial$ is the initial-state Bloch sphere direction unit vector and $r$ is the purity. Second, we will assume that the purity is very small. Specifically, as will be shown later, our analysis is valid when $r \ll 1/\sqrt{n}.$ Third, we assume that a single parameter-independent preparatory unitary $\uprep$ is applied to the entire system of qubits. This produces a channel \emph{input state} $\rhoi:= \uprep \rhoo^{\otimes n} \uprep^\dagger$ where $\rhoo^{\otimes n}$ is the initial state for the system of all $n$ qubits. Fourth, we assume that after the preparatory unitary the channel $\hat{\Gamma}(\lambda)$ is invoked once on a single qubit, mapping the channel input state to a final pre-measurement state via $\rhoi \stackrel{\hat{\Gamma}(\lambda)}{\mapsto} \rhof(\lambda)$. The entire process is illustrated in Fig~\ref{fig:singlechannelschemes}b).  
We term this a \emph{correlated-state} protocol since for low enough purity the states may be separable and the presence of entanglement is not assured (for details of separability in previously studied specific cases see~\cite{collins13,collins15}).

We will compare such a correlated-state protocol to one in which there is only a single qubit available and the channel is invoked once, as illustrated in Fig~\ref{fig:singlechannelschemes}a); we call the the latter the single-qubit, single-channel (SQSC) protocol. This yields the same QFI per channel invocation as an independent-channel protocol and thus in terms of the QFI per channel invocation, we are effectively comparing a correlated-state protocol to an independent-channel protocol. Therefore we ask whether, for given initial-state purity, there is an entangling preparatory unitary so that the correlated-state protocol of Fig.~\ref{fig:singlechannelschemes}b) yields a QFI exceeding than that of any SQSC protocol, for the same initial-state purity, and, to lowest non-trivial order in the initial-state purity, what gains such an correlated-state protocol provides.  

One physical motivation for this would be estimation of parameters associated with the evolution of single spins in room-temperature, solution-state NMR. Each molecule contains $n$ distinct nuclear spins; inter-molecular interactions average to zero and can be ignored. The entire ensemble only serves to amplify measurement signals and provide representative sampling of measurement outcomes. Therefore preparatory unitaries, channel actions and measurements may be regarded as restricted to within individual molecules.  Typically~\cite{nielsen00},  $r \approx 10^{-4}$ and thus our analysis applies whenever the number of nuclear spins within one molecule would be much less than $10^8.$ Most current room-temperature solution-state NMR lies well within this realm. 

Ultimately it may be of interest to compare correlated-state protocols in which the channel is invoked more than once. However, is known that for the particular cases of the phase-flip~\cite{collins13} and depolarizing channels~\cite{collins15} such correlated-state protocols do not yield advantages for all parameter values when there are two or more channel invocations. On the other hand, when there is only one channel invocation, these protocols definitely are advantageous over all parameter values for the phase-flip channel~\cite{collins13} and are probably so for the depolarizing channel~\cite{collins15}.  Thus, we only consider the situation where the channel is invoked once and assess whether the remaining ancillary qubits assist in the parameter estimation.

Sometimes low purity situations can be assessed using expressions for the QFI that are valid for all purities~\cite{modi11,collins13,collins15}.  Such exact expressions do not appear to be available for all channels and protocols. We will present a series approach for computing the QFI that is correct to the lowest non-trivial orders in the purity.

Before doing this, we consider what the framework by Escher et.al.~\cite{escher11} for noisy quantum metrology describes for this situation. The Escher framework considers a system in an initial \emph{pure} state $\rhoo$. The channel plus noise acts on this to produce a final pre-measurement state, which can be expressed~\cite{nielsen00} using a set of Kraus operators, $\left\{ \hat{\Pi}_l\right\}$, as $\rhof(\lambda) = \sum_l \hat{\Pi}_l(\lambda)\rhoo  \hat{\Pi}_l^\dagger(\lambda)$. Then the QFI is bounded~\cite{escher11} according to 
\begin{equation}
   H \leqslant 4 \left[ \Trace{\left(\hat{H}_1 \rhoo\right)} - \left( \Trace{\left(\hat{H}_2 \rhoo\right)} \right)^2\right]
   \label{eq:escherbound}
\end{equation}
where $\hat{H}_1 := \sum_l \frac{\partial \hat{\Pi}_l^\dagger}{\partial \lambda} \frac{\partial \hat{\Pi}_l}{\partial \lambda}$ and $\hat{H}_2 := i \sum_l \frac{\partial \hat{\Pi}_l^\dagger}{\partial \lambda} \hat{\Pi}_l.$ For any given channel plus noise there are multiple Kraus representations and that which minimizes the right hand side of Eq.~\eqref{eq:escherbound} exactly yields the QFI~\cite{escher11}. The Escher framework does not give an explicit method for finding the optimal Kraus representation. Sometimes careful choices of Kraus operators yield useful bounds on the QFI~\cite{escher11}. This framework can address single-qubit channel parameter estimation with noisy initial states by regarding the input state for each qubit as produced from a pure state via a depolarizing channel. This gives Kraus operators that combine the channel and initial depolarization Kraus operators and these can be used to compute the right hand side of Eq.~\eqref{eq:escherbound}. As shown in appendix~\ref{app:escherbound}, applying this to the example of a phase-flip channel acting on a single qubit gives a bound that is in excess of the known QFI for this channel acting on a noisy input state; in fact the bound does not even refer to the initial-state purity. This is a consequence of the choice of Kraus representation and there must clearly be a more restrictive choice, but what that is is not apparent and therefore the Escher framework is not immediately instructive in cases where the noise appears in initial states. 


\subsection{Series computation of the QFI}

If the initial-state purity is sufficiently low then we can approximate the QFI to lowest order in the purity by expressing the QFI and its constituent ingredients as series in increasing powers of the purity; this circumvents difficulties associated with exact computation of the SLD.  

To do so, the initial state can be expressed as 
\begin{equation}
 \rhoo^{\otimes n} = \sum_{j=0}^n r^j\; \rhooorder{j},
\label{eq:initialall}
\end{equation}
where $\rhooorder{j}$ is an operator independent of the purity. The preparatory unitary maps this to the input state $\rhoi:= \uprep \rhoo^{\otimes n} \uprep^\dagger,$ and thus 
\begin{equation}
 \rhoi = \sum_{j=0}^n r^j\; \rhoiorder{j}
\end{equation}
where $\rhoiorder{j} = \uprep^\dagger \rhooorder{j} \uprep$ is again independent of $r$. Similarly the final state can be expressed as
\begin{equation}
 \rhof(\lambda) = \sum_{j=0}^n r^j\; \rhoforder{j}(\lambda)
\label{eq:finalall}
\end{equation}
where $\rhoforder{j}(\lambda)$ is completely determined by evaluating the channel actions on $\rhoiorder{j}.$

Given that the state for each qubit is $\rhoo = (\iop + r \sigmagen{\rinitial} )/2$, two lowest order initial-state terms for the $n$ qubit system are 
\begin{subequations}
\begin{eqnarray}
  \rhooorder{0} & = & \frac{1}{N}\; \iop^{\otimes n} \quad \textrm{and}\\
  \rhooorder{1} & = & \frac{1}{N}\; \left[ 
	                                    \sigmagen{\rinitial} \otimes \iop^{\otimes (n-1)} 
																			+ \iop \otimes \sigmagen{\rinitial}  \otimes \iop^{\otimes (n-2)} 
																		\right. \nonumber \\
                 &  & \phantom{\frac{1}{N}\;} \left.
																			+ \cdots 
																			+  \iop^{\otimes (n-1)} \otimes \sigmagen{\rinitial} 
																		\right]
\end{eqnarray} 
\label{eq:initialrhoorders}
\end{subequations}
where $N = 2^n.$  For any preparatory unitary 
$
 \rhoiorder{0}  =  \iop^{\otimes n}/N
$
since $\uprep \iop \uprep^\dagger = \iop$. Higher order terms in $\rhoi$ depend on the preparatory unitary. There are no simple general expressions for the lowest order terms in the final state as certain channels, such as the amplitude damping channel map the identity in a non-trivial way.  

Similarly the SLD and QFI can be expressed as power series, possibly with infinitely many terms, in $r$. Thus  
\begin{equation}
 \score(\lambda) = \sum_{j=0}^\infty r^j\; \scoreorder{j}(\lambda),
\label{eq:scoreall}
\end{equation}
where $\scoreorder{j}(\lambda)$ is an operator independent of $r$,  and 
\begin{equation}
 H = \sum_{j=0}^\infty r^j\; \Horder{j},
\label{eq:hall}
\end{equation} 
where $\Horder{j}$ is independent of $r$. The operators $\scoreorder{j}(\lambda)$ can be evaluated by substituting from Eqs.~\eqref{eq:finalall} and~\eqref{eq:scoreall} into Eq.~\eqref{eq:slddefintion}. The result must be true for all values of $r$ and comparing terms order by order gives
\begin{equation}
 \frac{\partial \rhoforder{k}}{\partial \lambda} = \frac{1}{2}\;
                                                         \sum_{j=0}^k \left( \scoreorder{k-j} \rhoforder{j} 
																															              + \rhoforder{j} \scoreorder{k-j}
																															             \right).
\label{eq:sldorders}
\end{equation}

Similarly substituting from Eqs.~\eqref{eq:finalall} and \eqref{eq:scoreall} into Eq~\eqref{eq:quantumfishertwo} gives
\begin{equation}
  \Horder{j} = \sum_{k=0}^j \Trace{\left[ \frac{\partial \rhoforder{j-k}}{\partial \lambda} \scoreorder{k} \right]}.
\label{eq:horders}
\end{equation}
This allows for an iterative calculation of the QFI in increasing orders of the purity parameter; for sufficiently low purities, the QFI can be approximated by truncation. 

It is also useful to determine series expressions for the eigenstates of the SLD in order to assess possible measurements that  saturate the quantum CRB. Denote the normalized eigenstate of the SLD by $\ket{\phi}$ and the associated eigenvalue by $\mu$. Again these can be expanded as power series in $r$, giving 
\begin{equation}
  \ket{\phi} = \sum_{j=0}^\infty r^j \scorestatej
\end{equation}
and
\begin{equation}
  \mu = \sum_{j=0}^\infty r^j \scoreevalj.
\end{equation}
The normalization condition $\innerprod{\phi}{\phi} = 1$ must hold for all $r$ and implies that $\innerprod{\phi^{(0)}}{\phi^{(0)}} =1.$ Then order-by-order comparison of terms in $\score \ket{\phi} = \mu \ket{\phi}$ gives, for each $k=0,1,\ldots,$
\begin{equation}
  \sum_{j=0}^k \scoreorder{k-j}\scorestatej = \sum_{j=0}^k \scoreevalkj\scorestatej.
\label{eq:scoreeigenorders}
\end{equation}
This yields an iterative scheme for determining the eigenstates of the SLD and hence one possible projective measurement which saturates the quantum CRB. 


\subsection{Unital channels}

A unital channel maps $\iop \stackrel{\Gamma}{\rightarrow} \iop$ and here, $\rhoforder{0} = \iop^{\otimes n}/N$. Repeatedly using  Eq.~\eqref{eq:sldorders} results in
\begin{subequations}
\begin{eqnarray}
  \scoreorder{0} & = & 0 \quad \textrm{and} \\
	\scoreorder{1} & = & N \frac{\partial \rhoforder{1}}{\partial \lambda}.
\end{eqnarray}
\label{eq:unitalsld}
\end{subequations}
Thus, for unital channels,  Eq.~\eqref{eq:horders} yields 
\begin{subequations}
\begin{eqnarray}
  \Horder{0} & = & 0,\\
	\Horder{1} & = & 0, \quad \textrm{and} \\
	\Horder{2} & = & N \Trace{\left[ \left( \frac{\partial \rhoforder{1}}{\partial \lambda} \right)^2\right] }.
\end{eqnarray}
\label{eq:unitalqfi}
\end{subequations}

This immediately establishes the result, found earlier~\cite{modi11, collins13, collins15} for the qubit phase-shift, phase-flip and depolarizing channels, that the lowest order terms for the QFI are second order in the purity, provided that the preparation step consists of unitary operations only.  

Additionally Eq~\eqref{eq:scoreeigenorders} allows for computation of the eigenstates of the SLD via
\begin{subequations}
\begin{eqnarray}
  \scoreorder{0} \scorestatezero & = & \scoreevalzero \scorestatezero \quad \textrm{and} \\
  \scoreorder{1} \scorestatezero + \scoreorder{0} \scorestateone & = & \scoreevalzero \scorestateone + \scoreevalone \scorestatezero.
\label{eq:unitaleigenstates}
\end{eqnarray}
\end{subequations}
Since $\scorestatezero \neq 0,$ but $\scoreorder{0} = 0$, the first gives $\scoreevalzero =0$, leaving 
\begin{equation}
  \scoreorder{1} \scorestatezero  = \scoreevalone \scorestatezero = \frac{\partial \rhoforder{1}}{\partial \lambda} \scorestatezero.
\end{equation}
Thus, to lowest order in the purity, a measurement that suffices to saturate the quantum CRB is one which is done in the eigenbasis of $\frac{\partial \rhoforder{1}}{\partial \lambda}$.


\section{Single-qubit, single-channel protocols}
\label{sec:sqsc}

A baseline against which to compare any metrology protocol is the SQSC protocol illustrated in Fig.~\ref{fig:singlechannelschemes}a). The analysis depends on whether the channel is unital nor not and the results will be described in terms of the Bloch-sphere mapping of Eq.~\eqref{eq:Blochspherevectorevol}. 

\subsection{SQSC protocols for unital channels}

For a single qubit unital channel $\rhoiorder{1}: = \sigmagen{\rinitial}/2$, giving $\rhoforder{1} = \left( M\rinitial \right) \cdot \hat{\myvector{\sigma}}/2.$  This and Eqs.~\eqref{eq:unitalsld} and \eqref{eq:unitalqfi} imply that, \emph{to lowest order in the purity,}
\begin{equation}
 H = r^2\; \rinitial^\top \dot{M}^\top \dot{M} \rinitial 
\label{eq:sqscunital}
\end{equation}
where the dot indicates the derivative with respect to the parameter. This forms a general result for SQSC protocols for unital channels. 

Further analysis, all to lowest order only in the purity, uses the singular value decomposition for real matrices. Here $\dot{M} = A S B$ where $A$ and $B$ are each orthogonal $3\times3$ matrices and $S = s_1 P_1 + s_2 P_2 + s_3 P_3$ is a diagonal matrix with positive entries arranged so that $s_1 \geqslant s_2 \geqslant s_3$; here $\{P_i\}$ are projectors onto each of the three orthogonal directions associated with unit vectors $\{ \unitvec{e}_1, \unitvec{e}_2, \unitvec{e}_3 \}$. The orthogonality of $A$ and projective nature of $P_i$ implies that
\begin{equation}
 H = r^2\; \sum_{i=1}^{3} s_i^2 \rinitial^\top B^\top P_i B \rinitial.
\label{eq:sqscunitalsingvalues}
\end{equation}
Now $\rinitial^\top B^\top P_i B \rinitial \geqslant 0$ and $\sum_{i=1}^{3} \rinitial^\top B^\top P_i B \rinitial = 1$ implies that the \emph{optimal lowest order SQSC protocol QFI} is
\begin{equation}
 \Hsopt = r^2\; s_1^2.
 \label{eq:hsqscunitalopt}
\end{equation}
This is attained with $\rinitial = B^\top \unitvec{e}_1$ where $\unitvec{e}_1$ is the unit vector associated with the maximum singular value in $S$. Note that, depending on the singular value decomposition this might depend on the parameter to be estimated. 

One measurement which can saturate the quantum CRB bound is a projective measurement onto the eigenbasis of $\frac{\partial \rhoforder{1}}{\partial \lambda}$. Here, for the optimal choice of input state,
%
 $\frac{\partial \rhoforder{1}}{\partial \lambda} = \left( \dot{M}  B^\top \unitvec{e}_1 \right) \cdot \hat{\myvector{\sigma}}/2
                                                 = \left( AS \unitvec{e}_1 \right) \cdot \hat{\myvector{\sigma}}/2$
%
and the resulting projective measurement operators are
\begin{equation}
 \hat{\Pi}_\pm := \frac{1}{2}\; \left[ 
                            \iop \pm \left( A\unitvec{e}_1 \right) \cdot \hat{\myvector{\sigma}}
                          \right].
\label{eq:SLDmeasurement}
\end{equation}
Whenever the direction of $A\unitvec{e}_1$ depends on the parameter,  these projectors will also depend on the parameter to be estimated and adaptive measurement schemes~\cite{barndorff00} must be invoked to attain the QFI. But if the direction of $A\unitvec{e}_1$ is independent of the parameter, then the method described here will yield a parameter-independent saturating measurement.

To summarize, with a unital channel subject to the SQSC protocol, the optimal QFI to lowest order in the purity is determined by finding the Bloch-sphere matrix $M$ that represents the channel action and  determining the singular value decomposition, $\dot{M} = A S B$. The optimal QFI depends only on the maximal singular value $s_1$ and the protocol which attains this is to prepare the input state along the Bloch sphere direction $B^\top \unitvec{e}_1$, where $\unitvec{e}_1$ is the direction associated with the maximal singular value in $S$, and then subject the qubit to the channel. One measurement that saturates the QCRB is a projection onto the Bloch-sphere direction $A\unitvec{e}_1.$ 

\emph{Example: Unitary phase shift.} The unitary phase shift about the $z$ axis through angle $\lambda$, is represented by $\rhoi \mapsto \rhof = U^\dagger\rhoi U$ where $U:= e^{-i\lambda \sigmaz/2}$. In the basis $\{ \unitvec{x}, \unitvec{y}, \unitvec{z}\},$
\begin{equation}
 M = \begin{pmatrix}
	    \cos{\lambda} & -\sin{\lambda} & 0 \\
	    \sin{\lambda} & \cos{\lambda} & 0 \\
	    0 & 0 & 1
     \end{pmatrix}.
\end{equation}
Then 
\begin{equation}
 \dot{M} = \begin{pmatrix}
	    -\sin{\lambda} & -\cos{\lambda} & 0 \\
	    \cos{\lambda} & -\sin{\lambda} & 0 \\
	    0 & 0 & 0
     \end{pmatrix}
\end{equation}
which gives $S=\diag{(1,1,0)}$ with various possibilities for $A$ and $B$. The vector associated with the maximal singular value is any unit vector in the the $xy$ plane. This gives an optimal lowest order QFI of $\Hsopt = r^2$. The optimal QFI is attained using a state with Bloch-sphere input direction in the $xy$ plane, for example $B^\top\unitvec{x}$. The saturating measurement of Eq~\eqref{eq:SLDmeasurement} is a projective measurement along the direction $A\unitvec{x}$. It is not possible that both the choice of initial Bloch-sphere direction and measurement direction can both be independent of the parameter; this is consistent with exact calculations. 

\emph{Example: Phase-flip channel.} The phase-flip channel maps $\rhoi \mapsto \rhof = (1-\lambda) \rhoi + \lambda \sigmaz \rhoi \sigmaz$. In the basis $\{ \unitvec{x}, \unitvec{y}, \unitvec{z}\},$
$M = \diag{(1-2\lambda, 1-2\lambda, 1)}$ and $\dot{M} = \diag{(-2,-2,0)}$ so that $S=\diag{(2,2,0)}$ with $A=\diag{(-1,-1,0)}$ and $B=I$ as one possibility. This gives an optimal lowest order QFI is $\Hsopt = 4 r^2$, attained when the initial-state Bloch-sphere direction is in the $xy$ plane. This agrees with approximations from the exact QFI~\cite{collins13}. The saturating measurement of Eq~\eqref{eq:SLDmeasurement} is a projection along the direction $\rinitial$ and is parameter independent.

\emph{Example: Depolarizing channel.} The depolarizing channel maps $\rhoi \mapsto \rhof = (1-\lambda) \Trace{[\rhoi]} \iop + \lambda \rhoi$ and $M = \lambda I$ with $\dot{M} = I.$ This indicates that the optimal lowest order QFI is $\Hsopt = r^2$ and this is attained regardless of the choice of initial-state vector. This is consistent with approximations for exact the QFI~\cite{collins15}. Again a saturating measurement from the SLD is parameter independent. 


\subsection{SQSC protocols for non-unital channels}

For the more general non-unital channel acting on a single qubit $\iop \stackrel{\Gamma}{\rightarrow} \iop + \sigmagen{d}$ and thus 
\begin{equation}
 \rhoforder{0} = \frac{1}{2}\; \left( \iop + \sigmagen{d} \right).
 \label{eq:unitalsqsizero}
\end{equation}
The resulting analysis, again all to lowest order in the purity, depends on whether $\myvector{d}$ is parameter dependent. In Appendix~\ref{app:sqscnonunital} we show that if $\myvector{d}$ is parameter dependent then the zeroth order term in the QFI is generally non-zero and is
\begin{equation}
 \Hsopt =  \begin{cases}
	                  \dot{\myvector{d}} \cdot \dot{\myvector{d}}  + \dfrac{1}{4(1-d^2)} \left[ \dfrac{\partial d^2}{\partial \lambda} \right]^2
										\quad \textrm{if $d \neq 1$ and}
										\\[2ex]
                    \dot{\myvector{d}} \cdot \dot{\myvector{d}}
										\quad \textrm{if $d =1$.}
								\end{cases} 
 \label{eq:hsqscdynamicshiftopt}
\end{equation}
Here $d := \lvert\myvector{d}\rvert.$ If $\myvector{d}$ is parameter dependent then this is the optimal lowest order QFI for non-unital SQSC protocols. A key feature of such channels is that to lowest order in the purity, the QFI is independent of $r$ and this could be attained by an input state with zero purity. A sufficient measurement that would attain this is a projective measurement onto the eigenbasis of the lowest order score operator, $\scoreorder{0}$, and is thus a measurement along the Bloch sphere direction determined by $\dot{\myvector{d}}$ (if $d=1$) or $\dot{\myvector{d}}  + \frac{\partial \ln{(1-d^2)}}{\partial \lambda}\; \myvector{d}$ (if $d\neq1$). 

On the other hand, if $\myvector{d}$ is parameter independent then this will yield zero. Again as shown in Appendix~\ref{app:sqscnonunital}, if $\myvector{d}$ is independent of the parameter then the lowest order term in the QFI is 
\begin{eqnarray}
  H & = &   r^2\, \rinitial^\top \left[ 
	                   \dot{M}^\top \dot{M}
									 +
									 \frac{d^2}{1-d^2}\;
									 \dot{M}^\top  P_{\unitvec{d}}\dot{M} 
									 \right]
									 \rinitial,
 \label{eq:hsqscstaticshiftopt}
\end{eqnarray}
where $P_{\unitvec{d}}$ is the projector onto the direction $\unitvec{d}$. Note that if $d=1$ then $M=0$ and there is no parameter dependence to the channel at all. We can ignore this case. The entire operator within braces is positive and a singular value decomposition of this will eventually yield the optimal lowest order QFI and initial Bloch-sphere direction. Note that, comparing with Eq.~\eqref{eq:sqscunital},  this indicates that channels with nonzero constant Bloch-sphere shift vector will typically enhance the estimation accuracy of the channel corresponding to the Bloch-sphere matrix $M$ alone by effectively increasing the purity of the state.

\emph{Example: Generalized amplitude damping} The generalized amplitude damping channel maps $\rhoi \mapsto \rhof = \sum_{i=1}^4 E^\dagger_i \rhoi E_i$ where
\begin{eqnarray}
   E_1 & = & \sqrt{p}
	           \begin{pmatrix}
		           1 & 0 \\
							 0 & \sqrt{1-\lambda}
	           \end{pmatrix}
						 \nonumber \\
   E_2 & = & \sqrt{p}
	           \begin{pmatrix}
		           0 & \sqrt{\lambda}\\
							 0 & 0
	           \end{pmatrix}
						 \nonumber \\
   E_3 & = & \sqrt{1-p}
	           \begin{pmatrix}
		           \sqrt{1-\lambda} & 0 \\
							 0 & 1
	           \end{pmatrix}
						 \nonumber \\
   E_4 & = & \sqrt{1-p}
	           \begin{pmatrix}
		           0 & 0 \\
							 \sqrt{\lambda} &0
	           \end{pmatrix}
\end{eqnarray}
with $0 \leqslant p \leqslant 1.$

Then~\cite{nielsen00}, $M= \diag{(\sqrt{1-\lambda},\sqrt{1-\lambda}, 1-\lambda)}$ and $\myvector{d} = \lambda (2p-1) \unitvec{z}.$ This yields
%
 $\Hsopt = 1/[1-\lambda^2(2p-1)^2].$
%
The optimal measurement that saturates the quantum CRB bound is a projective measurement along $\unitvec{z}.$

Compiling these results gives in a complete characterization of the lowest order QFI terms for SQSC protocols for all channels: Eq.~\eqref{eq:hsqscunitalopt} for unital channels, Eq~\eqref{eq:hsqscdynamicshiftopt}  for non-unital channels with a parameter-dependent Bloch-sphere shift vector and  Eq~\eqref{eq:hsqscstaticshiftopt} for non-unital channels with a parameter-independent shift.


\section{Symmetric Pairwise Correlated Protocols}
\label{sec:corrstate}

The central question is whether there is an entanglement-assisted protocol which can yield a larger QFI per channel invocation than the optimal SQSC protocol with the same purity. Previous results for parameter estimation for the phase-shift, phase-flip and depolarizing channel showed that this is possible for a particular correlating preparatory unitary~\cite{modi11,collins13,collins15}. We consider a generalization of this for any qubit channel. 

\begin{figure}%
 \includegraphics[scale=0.6]{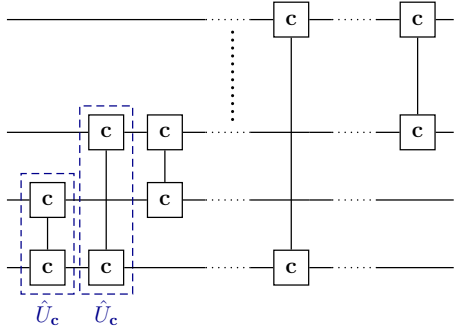}
\caption{The preparatory unitary for a general symmetric pairwise correlated scheme considered in this article. The symbols within the blue dashed frame represent a single iteration of $\uc$; the boxes indicate the two qubits on which the gate acts.  
				\label{fig:corrscheme}%
				}
\end{figure}

Specifically, we consider a protocol where the preparatory unitary is constructed from the two qubit unitary
\begin{equation}
  \uc := \frac{1}{2} \left( 
	               \iop \otimes \iop + 
	               \iop \otimes \sigmagen{c} +
								 \sigmagen{c} \otimes \iop - 
								 \sigmagen{c} \otimes \sigmagen{c}
								 \right) 
\end{equation}
where $\myvector{c}$ is a unit vector, which determines the \emph{Bloch-sphere control direction} of this gate. The preparatory unitary is defined to consist of a product of such unitaries, one for each distinct pair of qubits, as illustrated in Fig.~\ref{fig:corrscheme}. This is symmetrical under interchange of qubits and only involves pairwise correlating unitaries; we term it a symmetric pairwise correlated protocol. If $\myvector{c} = \unitvec{z}$ this is the controlled-Z gate used along with other single qubit Hadamard gates in the correlated-state protocols studied previously~\cite{collins13,collins15}.

Aside from demonstrating gains in the past, protocols of this type are interesting because the number of basic two qubit gates scales quadratically in the total number of qubits and in many physical settings these gates are relatively easily constructed. For example, in solution-state NMR implementations of quantum information processing they have been implemented experimentally since the outset of that field~\cite{cory00,ramanathan04,jones11}. 

Under this preparatory unitary, the two lowest order terms in the input state to the channels are 
$
 \rhoiorder{0} = \iop^{\otimes n}/N
$
and $ \rhoiorder{1} = \uprep \rhooorder{1} \uprep^\dagger$; these will be sufficient for determining the lowest order terms in the QFI. In Appendix~\ref{app:inputdensityop} we show that 
\begin{eqnarray}
 \rhoiorder{1} & = &
                \frac{1}{N}\;
								\left(
								  \sigmagen{\rinitial} \otimes \sigmagen{c}^{\otimes(n-1)}
									+ \cdots + 
									\sigmagen{c}^{\otimes(n-1)} \otimes \sigmagen{\rinitial}
								\right) 
								\nonumber \\
							& \phantom{=} & 
							  +
                \frac{\rinitial \cdot \myvector{c}}{N}\;
								\left(
								  \sigmagen{c} \otimes \iop^{\otimes(n-1)}
									+ \cdots + 
								  \iop^{\otimes(n-1)} \otimes \sigmagen{c}
								\right) 
								\nonumber \\
							& \phantom{=} & 
							  - 
								\frac{n\, \rinitial \cdot \myvector{c}}{N}\; 
								\sigmagen{c}^{\otimes n}.							 
\label{eq:corrorderoneinput}
\end{eqnarray}
Note that within the first parentheses, there are $n$ different terms, each containing a single factor of $\sigmagen{\rinitial}$. Similarly within the second parentheses there are also $n$ different terms, each containing a single factor of $\sigmagen{c}.$

We assume that the channel is invoked once on a single qubit. Again the analysis depends on whether the channel is unital or not.


\subsection{Symmetric pairwise correlated protocol for unital channels with a single invocation}

Assume that a unital channel acts once on the leftmost qubit in the tensor product representation. Then the terms in the first order term in the input state of Eq.~\eqref{eq:corrorderoneinput} are mapped by the channel as
%
\begin{eqnarray}
  \sigmagen{\rinitial} \otimes \sigmagen{c}^{\otimes(n-1)}
	& \mapsto &
	(M \rinitial) \cdot \hat{\myvector{\sigma}} \otimes \sigmagen{c}^{\otimes(n-1)}
	\nonumber
	\\
  \sigmagen{c} \otimes \sigmagen{\rinitial} \otimes \sigmagen{c}^{\otimes(n-2)}
	& \mapsto &
	(M \myvector{c}) \cdot \hat{\myvector{\sigma}} \otimes \sigmagen{\rinitial} \otimes \sigmagen{c}^{\otimes(n-2)}
	\nonumber 
	\\
  \sigmagen{c} \otimes \iop^{\otimes(n-1)}
	& \mapsto &
	(M \myvector{c}) \cdot \hat{\myvector{\sigma}} \otimes \iop^{\otimes(n-1)}
	\nonumber 
	\\
  \iop \otimes \sigmagen{c}   \otimes \iop^{\otimes(n-2)} 
	& \mapsto &
	\iop \otimes \sigmagen{c}   \otimes \iop^{\otimes(n-2)}
	\nonumber
	\\
	\sigmagen{c}^{\otimes(n)} 
	& \mapsto &
	(M \myvector{c})\cdot \hat{\myvector{\sigma}} \otimes \sigmagen{c}^{\otimes(n-1)} 
\end{eqnarray}
%
where $M$ is the channel Bloch-sphere matrix. Thus
\begin{eqnarray}
 \rhofderivorder{1} 
		& = &
		\frac{1}{N}\;
								\bigl[
								  (\dot{M}\rinitial) \cdot \hat{\myvector{\sigma}} \otimes \sigmagen{c}^{\otimes(n-1)}
								\bigr. 
		\nonumber \\
		&  & 
		\phantom{\frac{1}{N}\;}
									+ 
									(\dot{M} \myvector{c}) \cdot \hat{\myvector{\sigma}} \otimes \sigmagen{\rinitial} \otimes \sigmagen{c}^{\otimes(n-2)} 
		\nonumber \\
		&  & 
		\phantom{\frac{1}{N}\;}
									+		\cdots + 
								\bigl.
								(\dot{M} \myvector{c}) \cdot \hat{\myvector{\sigma}} \otimes \sigmagen{c}^{\otimes(n-2)} \otimes \sigmagen{\rinitial}
								\bigr]
		\nonumber \\
		&  & 
		+ \frac{\rinitial \cdot \myvector{c}}{N}\;
								(\dot{M} \myvector{c}) \cdot \hat{\myvector{\sigma}} \otimes \iop^{\otimes(n-1)}
		\nonumber \\
		&  &	
		- \frac{n \rinitial \cdot \myvector{c}}{N}\;
		 (\dot{M} \myvector{c}) \cdot \hat{\myvector{\sigma}} \otimes \sigmagen{c}^{\otimes(n-1)}.
\label{eq:rhofirstorderunital}
\end{eqnarray}
This yields our main result (see Appendix~\ref{app:corrqfi} for a proof) for unital channels: if the channel is invoked once on a single qubit when the symmetric pairwise correlated protocol is used then to the lowest order in the purity (remaining analysis is all to lowest order)
\begin{eqnarray}
 H & = & r^2 \rinitial^\top \left[ \left( I - \projc  \right)
                                        \dot{M}^\top \dot{M}
                                        \left( I - \projc  \right) 
																\right.
									\nonumber \\
						& & 
																\left.
						                            + (2-n) \projc   \dot{M}^\top \dot{M} \projc 
																	\right] \rinitial
									\nonumber \\
						& &		+ r^2 (n-1)\myvector{c}^\top \dot{M}^\top \dot{M}\myvector{c}
\label{eq:secondorderQFIcorr}
\end{eqnarray}
where $I$ is the $3\times3$ identity matrix and $\projc$ is the projector onto the control direction vector $\myvector{c}$. 

For a given channel there remains the task of choosing the control direction vector and initial-state Bloch-sphere vector so as to maximize the QFI of Eq.~\eqref{eq:secondorderQFIcorr}. The details depend on the channel but whenever a unital channel is invoked once it is possible to bound the lowest non-zero term in the QFI. As shown in Appendix~\ref{app:qfibound}, 
\begin{equation}
 n r^2 s_1^2 - r^2 s_1^2 \left( 1 -\frac{s_2^2}{s_1^2} \right) \leqslant H \leqslant n r^2 s_1^2,
 \label{eq:corrqfibounds}
\end{equation}
where $s_1 \geqslant s_2 \geqslant s_3 \geqslant 0$ are the singular values of $\dot{M}.$ The upper bound can only be saturated when $\rinitial$ and $\myvector{c}$ are perpendicular. The lower bound is attained for a particular choice of perpendicular $\rinitial$ and $\myvector{c}$ (see Appendix~\ref{app:qfibound} for details).

Equations~\eqref{eq:hsqscunitalopt} and~\eqref{eq:corrqfibounds} allow for comparison of the symmetric pairwise correlated protocol against the SQSC protocol to lowest order for unital channels. Here
\begin{equation}
  \left[ n - \left(1-\frac{s_2^2}{s_1^2}\right)\right]  \Hsopt \leqslant \Hcorropt \leqslant n \Hsopt
\end{equation}
where $\Hcorropt$ is the optimal QFI for the correlated protocol over all choices of $\myvector{c}$ and $\rinitial$. Thus
\begin{equation}
 n - \left(1- \frac{s_2^2}{s_1^2}\right)  \leqslant \frac{\Hcorropt}{\Hsopt} \leqslant n.
 \label{eq:corrgainseconorder}
\end{equation}
Since $s_2 \leqslant s_1$ this means that for large $n$ and to lowest order in the purity, the symmetric pairwise correlated protocol roughly gives an $n$-fold gain over the SQSC protocol for any unital channel.

Sometimes a precise statement about the optimal QFI for this correlated-state protocol can be made. If $s_1 = s_2$, as is true for several commonly considered channels, the the two bounds of Eq.~\eqref{eq:corrqfibounds} are identical and $\Hcorropt = n \Hsopt$. As another example, if $s_2=s_3=0$, the analysis of Appendix~\ref{app:directions} shows that $\Hcorropt= (n-1) \Hsopt$ and this is attained when $\myvector{c}$ and $\rinitial$ are perpendicular. 

The realm of applicability of the bounds of Eqs~\eqref{eq:corrqfibounds} and~\eqref{eq:corrgainseconorder} can be assessed via higher order terms in the QFI. In Appendix~\ref{app:higherorderqfiterms} we show that if $\rinitial$ and $\myvector{c}$ are perpendicular then $\Horder{3} =0$ and that generally
$\Horder{4}$ is of order $n^2.$ Thus the fourth order contribution to the QFI scales as $n^2r^4$. Given that the third order contribution is zero and that the second order contribution scales as $nr^2$ it is clear that approximating the QFI via the lowest order non-zero contribution is valid only when $nr^2 \ll 1.$ 


\subsection{Measurements for symmetric pairwise correlated protocol for unital channels}

The remaining issue with this optimal protocol is to find a QRB saturating measurement. A projective measure in the eigenbasis of $\rhofderivorder{1}$ suffices. For the optimal symmetric pairwise correlated protocol,  
\begin{eqnarray}
 \rhofderivorder{1} 
		& = &
		\frac{1}{N}\;
								\bigl[
								  (\dot{M}\rinitial) \cdot \hat{\myvector{\sigma}} \otimes \sigmagen{c}^{\otimes(n-1)}
								\bigr. 
		\nonumber \\
		&  & 
		\phantom{\frac{1}{N}\;}
									+ 
									(\dot{M} \myvector{c}) \cdot \hat{\myvector{\sigma}} \otimes \sigmagen{\rinitial} \otimes \sigmagen{c}^{\otimes(n-2)}
									+		\cdots
		\nonumber \\
		&  & 
		\phantom{\frac{1}{N}\;} + 
								\bigl.
								(\dot{M} \myvector{c}) \cdot \hat{\myvector{\sigma}} \otimes \sigmagen{c}^{\otimes(n-2)} \otimes \sigmagen{\rinitial}
								\bigr].
\label{eq:rhofirstorderunitalopt}
\end{eqnarray}
Sometimes this eigenbasis will depend on the parameter value, thus suggesting a measurement that would require knowledge of the parameter; this is a general issue which has been addressed elsewhere~\cite{barndorff00,toscano17}. Nonetheless the resulting measurement will always saturate the QFI.

\begin{figure}%
 \includegraphics[scale=0.6]{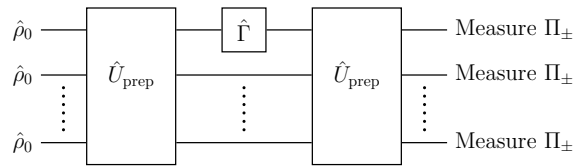}
\caption{Symmetric pairwise correlated scheme followed by local projective measurements. The measurements are chosen so that the associated projection operators are $\hat{\Pi}_\pm = \left( \iop \pm \sigmagen{\rinitial} \right)/2$ where $\rinitial$ is the initial state Bloch sphere direction.  
				\label{fig:corrschememeas}%
				}
\end{figure}

A separate issue is whether there exists a series of local, single qubit measurements that can yield the QFI. Whether such a procedure exists immediately after channel invocation is not clear. However, in Appendix~\ref{app:measurements} we show that if the channel invocation is following by another invocation of the preparatory unitary and this is followed by a local measurement on each qubit with appropriate choices of Bloch sphere directions, then the resulting classical Fisher information equals the lower bound of Eq.~\eqref{eq:corrqfibounds}. This merely entails another quadratic cost in terms of two-qubit gates. Figure~\ref{fig:corrschememeas} illustrates this scheme.

Specifically if, after the second invocation of the preparatory unitary, a local projective measurement corresponding to operators $\hat{\Pi}_\pm = \left( \iop \pm \sigmagen{\rinitial} \right)/2 $ is enacted on each qubit, where the Bloch sphere control and initial state vectors satisfy $\myvector{c} = B^\top \unitvec{e}_1$ and $\rinitial = B^\top \unitvec{e}_2$ where $\unitvec{e}_1$ and $\unitvec{e}_2$ are orthogonal unit vectors associated with the singular value decomposition of $\dot{M}$, then the classical Fisher information is
\begin{equation}
 F = n r^2 s_1^2 - r^2 s_1^2 \left( 1 -\frac{s_2^2}{s_1^2} \right).
\end{equation} 
This is exactly the lower bound of Eq.~\eqref{eq:corrqfibounds}.

Thus it is always possible to attain the gain with a factor of at least $n-1$ in estimation accuracy, using a local measurement scheme preceded by the preparatory unitary. We assess various important examples.

\emph{Example: Phase shift}
For the phase shift channel about the $z$ axis, $s_1=s_2=1,s_3=0$ and one possible choice of principal axis directions is $\unitvec{e}_1 = \unitvec{x}, \unitvec{e}_2 = \unitvec{y}$ and $\unitvec{e}_3 = \unitvec{z}$. By Eq.~\eqref{eq:corrqfibounds}, this gives an optimal QFI,
$ \Hcorropt = n \Hsopt.$ Taking $B=I$ in the singular value decomposition, the choices $\myvector{c} = \unitvec{e}_1$ and $\rinitial = \unitvec{e}_2$ attain this. Since $s_1=s_2$, the measurement scheme of Fig.~\ref{fig:corrschememeas} yields a classical Fisher information $F = n \Hsopt$ that saturates the optimal QFI with a parameter-independent measurement. 

\emph{Example: Phase flip}
Here $s_1=s_2=-2,s_3=0$ with $\unitvec{e}_3 = \unitvec{z}$ and $\unitvec{e}_1$ and $\unitvec{e}_2$ any perpendicular unit vectors in the $xy$ plane. This gives an optimal QFI, $\Hcorropt = n \Hsopt$. This is attained when $\myvector{c} = \unitvec{e}_1$ and $\rinitial=\unitvec{e}_2$.  Again $s_1=s_2$ and the measurement scheme of Fig.~\ref{fig:corrschememeas} yields a classical Fisher information $F = n \Hsopt$ with a parameter-independent choice of measurement. This agrees with lowest order approximations from exact expressions for all purities~\cite{collins13}.

\emph{Example: Depolarizing channel}
Here $s_1=s_2=s_3=1$, giving an optimal QFI,  an optimal QFI, $ \Hcorropt = n \Hsopt,$ which is attained when $\myvector{c} = \unitvec{e}_1$ and $\rinitial=\unitvec{e}_2$ are any two orthogonal vectors. Here also $s_1=s_2$ and the measurement scheme of Fig.~\ref{fig:corrschememeas} yields a classical Fisher information $F = n \Hsopt$ with a parameter-independent choice of measurement. This agrees with lowest order approximations from exact expressions for all purities~\cite{collins15}. 


\subsection{Gains for symmetric pairwise correlated protocol for unital channels}

These examples illustrate the general approximately $n$-fold gain offered by the symmetric pairwise correlated protocol whenever $n \ll 1/r^2.$ This is reminiscent of the gains described by the Heisenberg limit, where the QFI scales as $n^2$, over the standard quantum limit, where the QFI scales as $n$, and $n$ is the number of probes or channel invocations~\cite{giovannetti06}. One difference is that in the protocol of this article the channel only acts once on one of the probes whereas the typical $n$-fold Heisenberg scaling gain involves multiple copies of the channel or multiple systems subjected to the same channel. Mathematically the gain in our protocol arises from the fact that when the control and initial state Bloch sphere vectors are perpendicular the channel input state of Eq.~\eqref{eq:corrorderoneinput} has $n$ terms on which channel can have a non-trivial effect while, in the absence of the preparatory unitary, the initial state of Eq.~\eqref{eq:initialrhoorders} only offers one term on which the channel has a non-trivial effect. Somehow, the preparatory unitary of the symmetric pairwise correlated protocol has produced correlations that distribute information amongst the qubits so as to effectively mimic action of the channel as though it has acted on every qubit. It has also managed to do this in a way which works for all unital channels although the control and initial state Bloch sphere directions will need to be adjusted depending on the channel.  

Gains of this type in quantum metrology are often associated with the use of entangled states and it may be asked whether this is responsible for the gains here. However, as shown previously for the phase-shift~\cite{modi11}, phase-flip~\cite{collins13} and depolarizing channels~\cite{collins15}, for the two qubit case the system state are separable whenever the purity satisfies $r < \sqrt{2} -1$. This rules out entanglement as a source of the gains presented in this article. On the other hand, the same studies did show that for the two-qubit state, the quantum discord is non-zero for all purities, $r >0$, which encompass those of this article. How these statements might apply beyond two qubits is currently unclear.   


\subsection{Symmetric pairwise correlated protocol for non-unital channels}

For non-unital channels where $\myvector{d}$ depends on the parameter, the lowest non-zero term in the QFI is the zeroth order term. The zeroth order term in the density operator in the symmetric pairwise correlated-state protocol will be the same as that in the SQSC protocol; this is evident from setting $r=0$ in the formalism. Thus for such channels, the lowest order term in QFI in the symmetric pairwise correlated protocol is the same as that for the SQSC protocol. To lowest order there is nothing to gain from the correlated-state protocol for such cases. 

A possible explanation for this is that, for such non-unital channels, even the maximally mixed state would be fruitful for estimation as the channel maps, $\iop \stackrel{\Gamma}{\mapsto} \iop + \sigmagen{d}$ and the parameter could be estimated from the final system state. A maximally mixed initial state would be unaffected by any preparatory unitary and thus any correlated-state protocol would not make a difference for estimation if this were the system's initial state. The lowest order analysis that we have used only retains this maximally mixed term and therefore we might not have expected any gains from the correlated-state protocol here.


\section{Summary of Results}
\label{sec:summary}

\begin{figure*}%
 \includegraphics[scale=0.85]{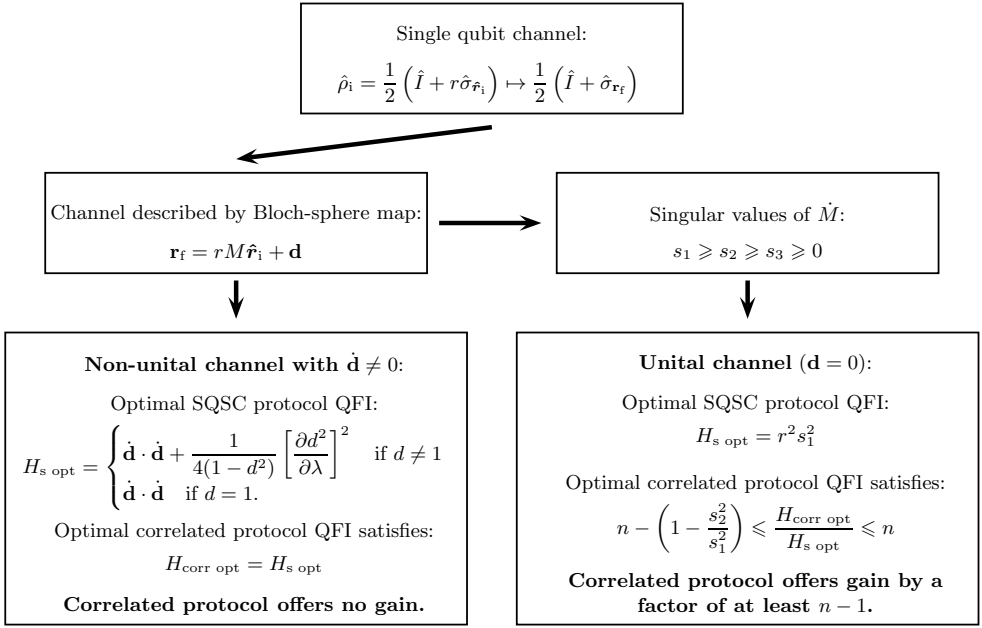}
\caption{Summary of the key results for single qubit channel parameter estimation. Results are all to lowest non-zero order in purity. The correlated protocol refers  to a symmetric  pairwise correlated-state protocol with the channel invoked once on one of $n$ qubits. 
				\label{fig:summary}%
				}
\end{figure*}

We summarize the main results for the single-qubit single-channel (SQSC) protocol and the symmetric pairwise correlated protocols (see Figs~\ref{fig:singlechannelschemes} and~\ref{fig:corrscheme} and the associated descriptions). These assume that all qubits are initially in the state $\rhoo = (\iop + r \sigmagen{\rinitial})/2$ and the results apply whenever the purity satisfies $nr^2 \ll 1$. Figure~\ref{fig:summary} provides a schematic that describes the main results in terms of the relevant quantum Fisher information for single qubit channels. 
%


\section{Conclusion}
\label{sec:discussion}

We have compared quantum parameter estimation protocols for qubit channels when the available states are mixed with very low purity and where the channel is invoked once. We have shown that for any unital channel, with initial-state purity $r$, the particular $n$ qubit correlated input state generated by the symmetric pairwise correlated protocol provides a roughly $n$-fold increase in estimation accuracy over protocols that use uncorrelated states provided that $nr^2 \ll 1$. These results agree with approximations from exact results for the known special cases of the phase-shift~\cite{modi11}, phase-flip~\cite{collins13} and depolarizing channels~\cite{collins15}. We also presented a measurement scheme that requires one more application of the preparatory unitary followed by local single qubit measurements that yields a classical Fisher information which saturates the lower bound in the QFI and this gives the roughly $n$-fold gain in accuracy. This is parameter-independent. There still remains the issue of finding generic measurement choices that are independent of the parameter and that yield a classical Fisher information which exactly saturates the optimal QFI. 

For non-unital channels with a parameter-dependent shift, to lowest order in the purity, there is no improvement in estimation accuracy using these particular parameters; this provides a first glimpse into amplitude-damping channel parameter with non-pure initial states. 

The formalism used here could be extended to situations where the channel is invoked on more than one of the qubits. The first order term in the channel input density operator would still have the form of Eq.~\eqref{eq:corrorderoneinput} but, for unital channels, the channel actions would produce an expression analogous to those of Eq~\eqref{eq:rhofirstorderunital} with more than one factor of the Bloch-sphere matrix appearing in each term. Differentiation would then yield expressions for the QFI involving both the Bloch-sphere matrix and its derivative. This would require modifying the singular value analysis used in this article. Additionally, it is already known for certain unital channels~\cite{collins13,collins15} that these protocols only offer advantages over a subset of possible parameter values and thus the universal results will not apply. Studies into restricted regions of the parameter space, where these protocols could offer advantages would still be warranted. 

The symmetric pairwise correlated-state protocol considered here generalizes those used previously and for unital channels. This is some improvement over previous studies~\cite{modi11,collins13,collins15} in estimation with noisy states, but even here the scheme awaits optimization over choices of initial-state Bloch-sphere and control vector directions. It is also possible that another type of preparation scheme might be optimal. The structure of the lowest order terms in the density operator in this formalism give some insight into the origin of the increase in the QFI. After the preparatory unitary, every qubit provides a term in the expression for the system state such that the channel has a non-trivial action on this term. This might be able to yield some insights into the origins of the accuracy enhancement and possible ways to improve it.

Finally, we note that there have been many studies of parameter estimation, typically of unitary parameters, in the presence of noisy processes (for example~\cite{huelga97,escher11,kolodynski13,jarzyna15}). Such studies consider the situation where noise appears during or after action of the channel whose parameter is to be estimated and usually yield bounds on the QFI that depend on noise parameters. For example, in~\cite{escher11} a general bound was developed and applied to optical phase estimation where noise was introduced during the phase evolution. The framework presented there resulted in a QFI which, when sufficient noise is present, scales as the number of probes, $n$, and does not yield an $n$-fold advantage over independent channel invocation protocols. However, although this study presents a general technique, these scaling results appear for the specific situation where the state space for the system is infinite dimensional.  Our initial attempts to use the general framework presented there for qubit systems where the noise is introduced at the outset have not yielded instructive bounds. In part, this must be due to the choice of representation of channel that also includes initial noise. However, the framework of~\cite{escher11} omits a specification of the optimal choice of such a channel. The exact relationship between this framework and the situation where noise is initially present in qubit channels warrants further investigation. 


\acknowledgments

The author would like to thank the Grup d'Informaci\'{o} Qu\`{a}ntica at the Universitat Aut\`{o}noma de Barcelona for hosting him while this study was conducted and Ramon Mu\~noz-Tapia for useful discussions.  


\appendix


\section{Channel parameterization}
\label{app:channelparameterization}

Consider a channel whose effect is described by Eq.~\eqref{eq:Blochspherevectorevol}. First, $\lvert \rfinal \rvert \leqslant 1$ for all possible inputs and for $r=0$ this implies $\rvert \myvector{d} \lvert \leqslant 1.$ We argue that if $M \neq 0$, then $\rvert \myvector{d} \lvert \neq 1.$ To do so, $\lvert \rfinal \rvert^2  = r^2 \rinput^\top M^\top M \rinput+ \rvert \myvector{d} \lvert^2 + r \rinput^\top M^\top\myvector{d} + r\myvector{d}^\top M \rinput$ where $\top$ indicates the transpose. Assuming that $\rvert \myvector{d} \lvert=1$, taking the case where $r=1$ and noting that the first term is positive this implies that for any unit vector $\rinput,$ one of the terms $\myvector{d}^\top M \rinput$ or $\rinput^\top M^\top\myvector{d} = \left( \myvector{d}^\top M \rinput \right)^\top$ must be negative. Thus for any unit vector $\rinput,$ $\myvector{d}^\top M \rinput$ must be negative. Now the singular value decomposition implies that $M=ASB$ where $A$ and $B$ are orthogonal matrices and $S$ is diagonal with positive entries along the diagonal. Thus for any $\rinput,$  $\myvector{d}^\top ASB \rinput$  must be negative. Letting $\myvector{d}^\prime := A^\top \myvector{d}$ and $\rinput^\prime := B \rinput$ gives that $\myvector{d}^\top ASB \rinput = \sum_j d^\prime_j s_j r_j^\prime$ where $s_j \geqslant 0$ are the singular values of $M$.  

 Thus, regardless $\myvector{d}^\prime$ there will always be some choices of $\rinput^\prime$ such that this is positive. This implies that, regardless of of $\myvector{d},$ there will be some choices of $\rinput$ so that $\myvector{d}^\top M \rinput$ is positive.  Thus, $\rvert \myvector{d} \lvert=1$ is only possible when $M=0.$ 


\section{Escher framework for phase flip channel parameter estimation with noisy initial states}
\label{app:escherbound}

We consider estimating the parameter $\lambda$ in a phase flip channel that acts on a single qubit according to $\rhoi \mapsto (1-\lambda) \rhoi + \lambda \sigmaz  \rhoi  \sigmaz$. The initial state for an individual qubits  $\rhoo = (\iop + r \sigmagen{\rinitial} )/2$  can be generated from a pure initial state $\rho_\mathrm{p} = (\iop + \sigmagen{\rinitial} )/2$ by a depolarizing channel with depolarizing parameter $r$. Denote the Kraus operators that generate the product of $n$ such mixed states from $n$ pure states by $\{ \hat{\Pi}_\textrm{d$j$}(r)\}$. The preparatory unitary $\uprep$ generates the channel input state $\rhoi = \sum_j \uprep \hat{\Pi}_\textrm{d$j$} \rho_\mathrm{p}^{\otimes n} \hat{\Pi}_\textrm{d $j$}^\dagger \uprep^\dagger$. Then denote the Kraus operators for the channel by $\{ \hat{\Pi}_\textrm{ch $k$}(\lambda)\}$. Thus the final pre-measurement state is $\rhof = \sum_{j,k} \hat{\Pi}_\textrm{ch $k$} \uprep \hat{\Pi}_\textrm{d$j$} \rho_\mathrm{p}^{\otimes n} \hat{\Pi}_\textrm{d$j$}^\dagger \uprep^\dagger \hat{\Pi}_\textrm{ch $k$}^\dagger$. Together these give \emph{one set} of Kraus operators $\{ \hat{\Pi}_\textrm{ch $k$}(\lambda) \uprep  \hat{\Pi}_\textrm{d$j$}(r)\}$  that map the pure initial state to the final state produced by the channel. It follows that the operators used in the Escher framework are
\begin{eqnarray}
   \hat{H}_1 & = & \sum_{j,k} \frac{\partial }{\partial \lambda} \left( \hat{\Pi}_\textrm{ch $k$}(\lambda) \uprep  \hat{\Pi}_\textrm{d$j$}(r) \right)^\dagger \nonumber \\
	                 &   &                  \frac{\partial }{\partial \lambda} \left( \hat{\Pi}_\textrm{ch $k$}(\lambda) \uprep  \hat{\Pi}_\textrm{d$j$}(r) \right) \nonumber \\
									& = & \sum_{j,k} \hat{\Pi}_\textrm{d$j$}^\dagger \uprep^\dagger 
									                          \frac{\partial  \hat{\Pi}_\textrm{ch $k$}^\dagger}{\partial \lambda} 
									                          \frac{\partial  \hat{\Pi}_\textrm{ch $k$}}{\partial \lambda}
																						\uprep  \hat{\Pi}_\textrm{d$j$}.
\label{eq:escherphaseone}
\end{eqnarray}
Then straightforward algebra gives
\begin{equation}
  \Trace{\left( \hat{H}_1 \rho_\mathrm{p} \right)} = \Trace{\left(  
	                                          \sum_k 
									                          \frac{\partial  \hat{\Pi}_\textrm{ch $k$}^\dagger}{\partial \lambda} 
									                          \frac{\partial  \hat{\Pi}_\textrm{ch $k$}}{\partial \lambda}
																						\rhoi
	                                                                                                        \right)}
\label{eq:eq:escherphaseonetrace}
\end{equation}
where $\rhoi$ is the channel input state generated by the preparatory unitary from the noisy initial states.  Similarly
\begin{equation}
  \Trace{\left( \hat{H}_2 \rho_\mathrm{p} \right)} = i \Trace{\left(  
	                                          \sum_k 
									                          \frac{\partial  \hat{\Pi}_\textrm{ch $k$}^\dagger}{\partial \lambda} 
									                          \hat{\Pi}_\textrm{ch $k$}
																						\rhoi
	                                                                                        \right)}.
\label{eq:eq:escherphasetwotrace}
\end{equation}
Now one possible Kraus representation for a single qubit phase flip channel is $\hat{\Pi}_\textrm{ch $1$} = \sqrt{1-\lambda}\, \iop$ and $\hat{\Pi}_\textrm{ch $2$} = \sqrt{\lambda}\, \sigmaz$. This yields, for a single qubit channel,
\begin{subequations}
\begin{eqnarray}
  \sum_k 
	    \frac{\partial  \hat{\Pi}_\textrm{ch $k$}^\dagger}{\partial \lambda} 
			\frac{\partial  \hat{\Pi}_\textrm{ch $k$}}{\partial \lambda}
	    & = & \frac{1}{4\lambda(1-\lambda)}\, \iop  \\		
  \sum_k 
	    \frac{\partial  \hat{\Pi}_\textrm{ch $k$}^\dagger}{\partial \lambda} 
			\hat{\Pi}_\textrm{ch $k$}
	    &= & 0.	
\end{eqnarray}																											
\label{eq:escherphaseops}
\end{subequations}
Consider now the situation where only one qubit is present and the phase-flip channel is invoked once. Then Eqs.~\eqref{eq:escherbound} and~\eqref{eq:escherphaseops} bound the QFI via $H  \leqslant 1/\lambda(1-\lambda)$ and this does not even refer to the initial-state purity.  However, the known optimal single qubit QFI~\cite{collins13} for the phase-flip channel acting on a single qubit with purity $r$ is $H = 4r^2/[1 - (1-2 \lambda)^2 r^2]$. Straightforward algebra shows that the bound given by the Escher framework and the particular choice of Kraus representation is strictly larger than the known optimal QFI whenever $r<1$.  


\section{QFI for SQSC protocols for non-unital channels}
\label{app:sqscnonunital}

Regardless of whether $\myvector{d}$ depends on the parameter or not, the lowest order version of Eq.~\eqref{eq:sldorders} gives
 $\rhofderivorder{0} = \left[ 
											        \scoreorder{0} \rhoforder{0} + \rhoforder{0} \scoreorder{0} 
											 \right]/2.$
The algebraic method offered in~\cite{collins13} gives
\begin{equation}
  \scoreorder{0} = 		\dfrac{1}{2}\; 
												\dfrac{\partial \ln{(1-d^2)}}{\partial \lambda}\; \iop 
												\\
												+ 
												 \left[
														 \dot{\myvector{d}}  
														 -  
														 \dfrac{1}{2}\;
														 \dfrac{\partial \ln{(1-d^2)}}{\partial \lambda}\; \myvector{d}
												 \right] \cdot \hat{\myvector{\sigma}}
\end{equation}
when $d := \lvert\myvector{d} \rvert \neq1$ and 
\begin{equation}
 \scoreorder{0} = \sigmagen{d}
\end{equation}
if $d= 1$.

Direct substitution into the lowest order version of Eq.~\eqref{eq:horders} yields the result of Eq.~\eqref{eq:hsqscdynamicshiftopt}.

If $\myvector{d}$ is parameter independent then $\rhofderivorder{0}=0$ and $\scoreorder{0} =0$ and, as before, $\Horder{0}=\Horder{1}=0.$ The next order term is attained via Eq.~\eqref{eq:sldorders}, which with Eq.~\eqref{eq:unitalsqsizero} gives
\begin{eqnarray}
 \rhofderivorder{1} & = & \frac{1}{2}\; \left[ \scoreorder{1} \rhoforder{0} + \rhoforder{0} \scoreorder{1} \right]
                          \nonumber \\
										& = & \frac{1}{2}\; \scoreorder{1} + \frac{1}{4}\; \left[ \scoreorder{1} \sigmagen{d} + \sigmagen{d} \scoreorder{1} \right]
\end{eqnarray}
If $d\neq 1,$ then the solution to this, which can be verified by direct substitution, is
\begin{eqnarray}
 \scoreorder{1} & = &
                  \frac{2-d^2}{1-d^2}\;
                  \rhofderivorder{2}
									- 
									\frac{1}{1-d^2}\;
									\left[
									 \rhofderivorder{1} \sigmagen{d}
									 + \sigmagen{d} \rhofderivorder{1}
									\right]
									\nonumber \\
								& & 
									+  
									\frac{1}{1-d^2}\;
									\left[
									 \sigmagen{d} \rhofderivorder{1} \sigmagen{d}
									\right].
 \label{eq:scoreorderonedindep}
\end{eqnarray}
The remaining case where $\myvector{d}$ is a parameter-independent unit vector requires $M=0$ and this leaves no parameter dependence. We can ignore this. Eqs.~\eqref{eq:hall} and~\eqref{eq:scoreorderonedindep} give that if $\myvector{d}$ is parameter independent, then to lowest order in the purity, 
\begin{eqnarray}
  H & = & r^2\, \frac{2 - d^2}{1-d^2}\; \Trace{\left[ \left( \rhofderivorder{1} \right)^2 \right]}
	                 \nonumber \\
						 &  &  -  r^2\, \frac{2}{1-d^2}\;
										   \Trace{\left[ 
											  \left( \rhofderivorder{1} \right)^2 \sigmagen{d} 
											 \right]}
	                 \nonumber \\
						 &  & +  r^2\, \frac{1}{1-d^2}\;
										   \Trace{
											 \left[
											   \sigmagen{d} \rhofderivorder{1} \sigmagen{d} \rhofderivorder{1}
											 \right]}. 
\end{eqnarray}
Then $\rhoforder{1} = \left( M\rinitial \right)\cdot \hat{\myvector{\sigma}}/2$ gives
\begin{eqnarray}
  H & = &   r^2\,  \rinitial^\top \dot{M}^\top \dot{M} \rinitial
									 +
									 \frac{r^2}{1-d^2}\;
									 \left( \myvector{d}^\top \dot{M} \rinitial \right)^2.
\end{eqnarray}
and this yields the result of Eq.~\eqref{eq:hsqscstaticshiftopt}. 


\section{Channel input states}
\label{app:inputdensityop}

Computing the effects of the preparatory unitary on the entails algebra of Pauli operators.  Explicit multiplication show that, for any vectors $\myvector{a}$ and $\myvector{b},$
\begin{equation}
  \sigmagen{a} \sigmagen{b}  = \myvector{a}\cdot\myvector{b} \iop 
	                           + i \left(\myvector{a}\times \myvector{b}\right) \cdot \hat{\myvector{\sigma}}
\label{eq:paulimulti}
\end{equation}
The effects of the preparatory unitary can be determined by via the action of $\hat{U}_{\myvector{c}}$ on pairwise products of operators. We will show that
\begin{eqnarray}
 \uc \left( \sigmagen{a} \otimes \iop \right) \ucconj & = & \sigmagen{a} \otimes \sigmagen{c} \nonumber \\
                                                      &   & + \left( \myvector{a}\cdot\myvector{c} \right)\,
																						                  \bigl( \sigmagen{c} \otimes \iop  - \sigmagen{c} \otimes \sigmagen{c} \bigr).
                                                             \label{eq:ucactionone} 
\end{eqnarray}

To demonstrate~\eqref{eq:ucactionone}, consider first $\left( \sigmagen{a} \otimes \iop \right) \ucconj$. Noting that $\ucconj = \uc$, repeatedly using~\eqref{eq:paulimulti} and the fact that $\myvector{c}$ is a unit vector gives
\begin{eqnarray}
   \left( \sigmagen{a} \otimes \iop \right) \ucconj & = & \frac{1}{2}\, 
	                                                                               \left[	 \myvector{a} \cdot \myvector{c}\,  
																																															 \iop \otimes \left( \iop - \sigmagen{c}\right)
																																													+ \sigmagen{a} \otimes \left( \iop + \sigmagen{c} \right)
																																								 \right. \nonumber \\
																																				&  &   \left. 
																																				          +i \sigmagen{a \times c} \otimes 
																																									    \left( \iop - \sigmagen{c}
																																									    \right)
																																								 \right].
\label{eq:ucactiononederivationone}
\end{eqnarray}
Then multiplying this separately by each term in $\uc$ gives, 
\begin{subequations}
\begin{eqnarray}
   \frac{1}{2}\,
   \iop \otimes \iop \left( \sigmagen{a} \otimes \iop \right) \ucconj & = & \frac{1}{4}\, 
	                                                                               \left[	 \myvector{a} \cdot \myvector{c}\,  
																																															 \iop \otimes \left( \iop - \sigmagen{c}\right)
																																								 \right. \nonumber \\
																																				&  &   \left. 
																																													+ \sigmagen{a} \otimes \left( \iop + \sigmagen{c} \right)
																																								 \right. \nonumber \\
																																				&  &   \left. 
																																				          +i \sigmagen{a \times c} \otimes 
																																									    \left( \iop - \sigmagen{c}
																																									    \right)
																																								 \right],
																																								 \label{eq:ucactiononederivationtwo} \\
   \frac{1}{2}\,
   \iop \otimes \sigmagen{c} \left( \sigmagen{a} \otimes \iop \right) \ucconj & = & \frac{1}{4}\, 
	                                                                               \left[	 - \myvector{a} \cdot \myvector{c}\,  
																																															 \iop \otimes \left( \iop - \sigmagen{c}\right)
																																								 \right. \nonumber \\
																																				&  &   \left. 
																																													+ \sigmagen{a} \otimes \left( \iop + \sigmagen{c} \right)
																																								 \right. \nonumber \\
																																				&  &   \left. 
																																				          -i \sigmagen{a \times c} \otimes 
																																									    \left( \iop - \sigmagen{c}
																																									    \right)
																																								 \right],
																																									\label{eq:ucactiononederivationthree} \\
   \frac{1}{2}\,
   \sigmagen{c} \otimes  \iop \left( \sigmagen{a} \otimes \iop \right) \ucconj & = & \frac{1}{4}\, 
	                                                                               \left[	\myvector{a} \cdot \myvector{c}\,
																																									        \left( \iop \otimes \iop 
																																													        + \iop \otimes \sigmagen{c} 
																																													\right.
																																								 \right. \nonumber \\
																																				&  &   \left. 
																																													\left.
																																																	+ 2 \sigmagen{c} \otimes \iop \
																																																	- 2 \sigmagen{c} \otimes \sigmagen{c}
																																													\right)
																																								 \right. \nonumber \\
																																				&  &   \left. 
																																													- \sigmagen{a} \otimes \left( \iop - \sigmagen{c} \right)
																																								 \right. \nonumber \\
																																				&  &     
																																								\negthickspace
																																								\negthickspace
																																								\negthickspace
																																								\left. 
																																				          i \sigmagen{c \times a} \otimes 
																																									    \left( \iop + \sigmagen{c}
																																									    \right)
																																								 \right], \textrm{and}
																																									\label{eq:ucactiononederivationfour} \\
   -\frac{1}{2}\,
   \sigmagen{c} \otimes  \sigmagen{c} \left( \sigmagen{a} \otimes \iop \right) \ucconj & = & \frac{1}{4}\, 
	                                                                               \left[	\myvector{a} \cdot \myvector{c}\,
																																									        \left( -\iop \otimes \iop 
																																													        - \iop \otimes \sigmagen{c} 
																																													\right.
																																								 \right. \nonumber \\
																																				&  &   \left. 
																																													\left.
																																																	+ 2 \sigmagen{c} \otimes \iop \
																																																	- 2 \sigmagen{c} \otimes \sigmagen{c}
																																													\right)
																																								 \right. \nonumber \\
																																				&  &   \left. 
																																													- \sigmagen{a} \otimes \left( \iop - \sigmagen{c} \right)
																																								 \right. \nonumber \\
																																				&  &   \left.
																																				          - i \sigmagen{c \times a} \otimes 
																																									    \left( \iop + \sigmagen{c}
																																									    \right)
																																								 \right].
\label{eq:ucactiononederivationfive}
\end{eqnarray}
\end{subequations}
Adding~\eqref{eq:ucactiononederivationtwo}-\eqref{eq:ucactiononederivationfive} gives~\eqref{eq:ucactionone}.

Now consider 
  $\rhoiorder{1} = \uprep \rhooorder{1} \uprep^\dagger$ where $\rhooorder{1} =\left[ 
	                                    \sigmagen{\rinitial} \otimes \iop^{\otimes (n-1)} 
																			+ \cdots 
																			+  \iop^{\otimes (n-1)} \otimes \sigmagen{\rinitial}
																		\right]/N$. 
Since the preparatory unitary is symmetric under interchange pair of qubits it suffices to evaluate 
$\uprep
   \sigmagen{\rinitial} \otimes \iop^{\otimes (n-1)}
	 \uprep^\dagger.
$
The $\uc$ factors in $\uprep$ and $\uprep^\dagger$ commute amongst each other and all of those that do not involve the leftmost qubit commute with the factors of the identity. Therefore, for each such pair of qubits a factor of $\uc$ multiplies a factor of $\ucconj$, leaving the identity. Thus we need only consider the factors of $\uc$ in $\uprep$ that involve the leftmost qubit.  This process can be illustrated with a three qubit system. Invoking~\eqref{eq:ucactionone} gives that the $\uc$ acting on the leftmost and center qubits produces
\begin{eqnarray}
  \sigmagen{\rinitial} \otimes \iop \otimes \iop
	& \mapsto &
	\sigmagen{\rinitial} \otimes \sigmagen{c} \otimes \iop
	\nonumber \\
	& & 
	  +\left(\rinitial \cdot \myvector{c}\right) 
	 \sigmagen{c} \otimes \iop \otimes \iop
	\nonumber \\
	& & 
	 - \left(\rinitial \cdot \myvector{c}\right)  \sigmagen{c} \otimes \sigmagen{c} \otimes \iop.
\label{eq:ucononeandtwo}
\end{eqnarray}
Now, noting that $\uc$ commutes with terms of the form $\sigmagen{c} \otimes \iop$ and $\sigmagen{c} \otimes \sigmagen{c}$, consider the factor $\uc$ acting on the leftmost and rightmost qubits. This leaves the second and third terms in Eq.~\eqref{eq:ucononeandtwo} unaltered since it commutes with them. However, according to~\eqref{eq:ucactionone} it acts on the first term of~\eqref{eq:ucononeandtwo} to produce
\begin{eqnarray}
  \sigmagen{\rinitial} \otimes  \sigmagen{c} \otimes \iop
	& \mapsto &	
	\sigmagen{\rinitial} \otimes  \sigmagen{c} \otimes  \sigmagen{c}
	\nonumber \\
	& & 
	+\left(\rinitial \cdot \myvector{c}\right)  \sigmagen{c}  \otimes  \sigmagen{c} \otimes \iop
	\nonumber \\
	& & 
	- \left(\rinitial \cdot \myvector{c}\right)   \sigmagen{c} \otimes  \sigmagen{c} \otimes  \sigmagen{c}.
\end{eqnarray}
Thus substituting this into the first term of~\eqref{eq:ucononeandtwo} reveals that under all factors of $\uc$ the terms $\sigmagen{c}  \otimes  \sigmagen{c} \otimes \iop$ cancel. Thus, under the preparatory unitary,
\begin{eqnarray}
  \sigmagen{\rinitial} \otimes \iop \otimes \iop
	& \mapsto &
	\sigmagen{\rinitial} \otimes \sigmagen{c} \otimes \sigmagen{c}
	 +
	\nonumber \\
	& &
	 + \left(\rinitial \cdot \myvector{c}\right)  \sigmagen{c} \otimes \iop \otimes \iop
	\nonumber \\
	& &
	- \left(\rinitial \cdot \myvector{c}\right) \sigmagen{c} \otimes \sigmagen{c} \otimes \sigmagen{c}.
\end{eqnarray}
This pattern continues and, for arbitrary numbers of qubits,
\begin{eqnarray}
  \uprep\;  
	\left( \sigmagen{\rinitial} \otimes \iop^{\otimes(n-1)}\right)\;  
	\uprep^\dagger
   & = &
		   \sigmagen{\rinitial} \otimes \sigmagen{c}^{\otimes(n-1)} \nonumber \\
	 & \phantom{=} & 
	     + \left( \rinitial \cdot \myvector{c} \right)\;  \sigmagen{c} \otimes \iop^{\otimes(n-1)} \nonumber \\
	 & \phantom{=} & 
			 - \left( \rinitial \cdot \myvector{c} \right)\; \sigmagen{c}^{\otimes n}.
\end{eqnarray}
This determines the term 
$\uprep
   \sigmagen{\rinitial} \otimes \iop^{\otimes (n-1)}
	 \uprep^\dagger.
$
This result, the symmetry of $\uprep$ under interchange of qubits, the form of $\rhooorder{1}$ and its symmetry under interchange of qubits then give the first order channel input term of~\eqref{eq:corrorderoneinput}.


\section{Lowest order symmetric pairwise correlated protocol QFI}
\label{app:corrqfi}

To prove the result of Eq.~\eqref{eq:secondorderQFIcorr}, note that it emerges from $H = r^2 \Horder{2} + {\cal O}(r^3)$ and the trace operation of Eq.~\eqref{eq:unitalqfi}, which requires computing the trace of the square of the entire right-hand side of Eq.~\eqref{eq:rhofirstorderunital}. This results in a sum of the trace of each term squared together with the traces of all ``cross terms''; we evaluate and list these separately. In both cases a useful tool is that 
%
 $\Trace{\left[ \sigmagen{a} \sigmagen{b} \right]} = 2\; \myvector{a}^\top \myvector{b}$
%
for any vectors $\myvector{a}$ and $\myvector{b}.$
Also note that the result will contain terms of the form $(\rinitial \cdot \myvector{c}) \dot{M}\myvector{c}$ and this can be expressed as $\dot{M}\projc\; \rinitial$ where $\projc$ is the projection operator onto $\myvector{c}$.

Then the squared terms of Eq.~\eqref{eq:rhofirstorderunital} are listed in Table~\ref{tab:squaredterms}.

\begin{table}[h]%
 \renewcommand{\arraystretch}{1.25}
\begin{tabular}{l|c|c}
 Term & Multiplicity & Contribution \\
 \hline
 1\textsuperscript{st} & $1$ & $\rinitial^\top \dot{M}^\top \dot{M}\rinitial/N$ \\
 \hline
 2\textsuperscript{nd} or 3\textsuperscript{rd} & $n-1$ & $\myvector{c}^\top \dot{M}^\top \dot{M}\myvector{c}/N$ \\
 \hline
 4\textsuperscript{th} & 1 & $\rinitial^\top \projc \dot{M}^\top \dot{M}\projc\rinitial/N$\\
 \hline
 5\textsuperscript{th} & 1 & $n^2\; \rinitial^\top \projc \dot{M}^\top \dot{M}\projc\rinitial/N$ \\
 \hline
\end{tabular}
\caption{Traces of square of terms from Eq.~\eqref{eq:rhofirstorderunital}. The multiplicity indicates the number of times each type of terms occurs in the product $\left( \rhofderivorder{1} \right)^2$. \label{tab:squaredterms}}
\end{table}

Similarly consider the ``cross terms'' of Eq.~\eqref{eq:rhofirstorderunital} are listed in Table~\ref{tab:crossterms}.

\begin{table}[h]%
 \renewcommand{\arraystretch}{1.25}
\begin{tabular}{l|c|c}
 Term & Multiplicity & Contribution \\
 \hline
 1\textsuperscript{st} and 2\textsuperscript{nd} or 3\textsuperscript{rd} & $2(n-1)$ & $\rinitial^\top \projc \dot{M}^\top \dot{M}\rinitial/N$ \\
 \hline
 1\textsuperscript{st} and 4\textsuperscript{th} & $1$ & $0$ \\
 \hline
 1\textsuperscript{st} and 5\textsuperscript{th} & $2$ & $-n\; \rinitial^\top \projc  \dot{M}^\top \dot{M}\rinitial/N$\\
 \hline
 2\textsuperscript{nd} and 3\textsuperscript{rd} & $(n-1)(n-2)$ & $\rinitial^\top \projc  \dot{M}^\top \dot{M} \projc\rinitial/N$ \\
 \hline
 2\textsuperscript{nd} or 3\textsuperscript{rd} and 4\textsuperscript{th} & $n-1$ & $0$ \\
 \hline
 2\textsuperscript{nd} or 3\textsuperscript{rd} and 5\textsuperscript{th} & $2(n-1)$ & $-n\; \rinitial^\top \projc  \dot{M}^\top \dot{M} \projc\rinitial/N$ \\
 \hline
 4\textsuperscript{th} and 5\textsuperscript{th} & $1$ & $0$ \\
 \hline
\end{tabular}
\caption{Traces of ``cross-term'' products from Eq.~\eqref{eq:rhofirstorderunital}. The multiplicity indicates the number of times each type of terms occurs in the product $\left( \rhofderivorder{1} \right)^2$.
         \label{tab:crossterms}}
\end{table}

Adding these gives
\begin{eqnarray}
  \Horder{2} & = & N \Trace{\left[ \left( \frac{\partial \rhoforder{1}}{\partial \lambda} \right)^2\right] }
	                \nonumber \\
						 & = & \rinitial^\top \left[ \dot{M}^\top \dot{M} 
						                            + (3-n) \projc  \dot{M}^\top \dot{M} \projc
																				-2 \projc  \dot{M}^\top \dot{M}
																	\right] \rinitial
	                \nonumber \\
							& & 	+(n-1)\myvector{c}^\top \dot{M}^\top \dot{M}\myvector{c}. 
\end{eqnarray}

Note that $\rinitial^\top \projc  \dot{M}^\top \dot{M}\rinitial = \rinitial^\top \dot{M}^\top \dot{M} \projc  \rinitial$ and thus a symmetric expression is
\begin{eqnarray}
 \Horder{2} &= &\rinitial^\top \left[ \dot{M}^\top \dot{M} 
						                            + (3-n) \projc  \dot{M}^\top \dot{M} \projc
														\right.
														\nonumber \\
						& & 
														\left.
																				- \projc  \dot{M}^\top \dot{M}
																				- \dot{M}^\top \dot{M} \projc 
																	\right] \rinitial
														\nonumber \\
						& & 
								+(n-1)\myvector{c}^\top \dot{M}^\top \dot{M}\myvector{c}.
\end{eqnarray}
Algebra then gives the stated result. 


\section{Bound on the lowest order term in the QFI for unital channels}
\label{app:qfibound}

The singular value decomposition is $\dot{M} = ASB,$ with $S = \sum_i s_i P_i$ arranged so that $s_1 \geqslant s_2 \geqslant s_3 \geqslant 0$ and where $P_i$ is a projector onto the unit vector $\unitvec{e}_i$ and each of these is one of $\unitvec{x},\unitvec{y}$ and $\unitvec{z}$. Then Eq.~\eqref{eq:secondorderQFIcorr} becomes
\begin{eqnarray}
 H & = & r^2 \sum_i s_i^2 \left\{ \rinitial^\top 
                                        \left( I - \projc \right) B^\top  P_i B
                                        \left( I - \projc  \right)
																			\rinitial
															 \right.
									\nonumber \\
						& &  \left.
													+ (n-1) \myvector{c}^\top P_i\myvector{c}
						                            - (n-2) \rinitial^\top \projc B^\top  P_i B \projc \rinitial 
																	\right\} .
\label{eq:secondorderQFIcorrsingvalues}
\end{eqnarray}

The lower bound of Eq.~\eqref{eq:corrqfibounds} can be established by the particular choice of $\myvector{c} = B^\top \unitvec{e}_1$ and $\rinitial = B^\top \unitvec{e}_2.$ The upper bound can be established by noting that, since $s_1^2 \geqslant s_2^2 \geqslant s_3^2 \geqslant 0$,
\begin{eqnarray}
 H & r^2 \leqslant & s_1^2 \sum_i \left\{ \rinitial^\top 
                                        \left( I - \projc  \right) B^\top  P_i B
                                        \left( I - \projc  \right)
																			\rinitial
															 \right.
									\nonumber \\
						& &  \left.
													+ (n-1) \myvector{c}^\top B^\top  P_i B \myvector{c}
															 \right.
									\nonumber \\
						& &  \left.
						                            - (n-2) \rinitial^\top \projc B^\top  P_i B \projc \rinitial 
																	\right\} .
\end{eqnarray}
Then the facts that $\sum_i P_i = I$, and $\myvector{c}$, $\rinitial$ are unit vectors and $\projc $, $I - \projc $ are projectors give
\begin{eqnarray}
 H & \leqslant & r^2 s_1^2 \left[ n - (n-1) \rinitial^\top \projc \rinitial 
																	\right].
\end{eqnarray}
The left side attains a maximum of $ns_1^2$ when $\myvector{c}$ and $\rinitial$ are perpendicular. This proves the result for the upper bound. It also implies that for the upper bound to be saturated the initial state Bloch-sphere vector direction and control direction must be perpendicular. But it does not guarantee that the upper bound can be saturated and, if not, it makes no statement about the directions of these vectors in order to attain the maximum QFI.


\section{Optimal QFI for $s_2 =s_3=0.$}
\label{app:directions}

To prove the result that the optimal choice of control and initial-state directions is one where they are perpendicular, consider any fixed choice of $\myvector{c}$. Then within Eq.~\eqref{eq:secondorderQFIcorrsingvalues} there appears the term 
$  W:=
   \rinitial^\top 
   \left( I - \projc \right) 
	 P_1
   \left( I - \projc \right)
	 \rinitial
	+ (n-1) \myvector{c}^\top P_1\myvector{c}
  - (n-2) \rinitial^\top \projc  P_1 \projc\rinitial
 $ 
and we will show that this is maximized when $\rinitial$ and $\myvector{c}$ are perpendicular. Note that $\myvector{c}$ and $\unitvec{e}_1$ span a plane. Then $\rinitial$ can be decomposed into a vector perpendicular to the plane $\myvector{r}_\textrm{0}^\perp$ and a vector parallel to the plane $\myvector{r}_\textrm{0}^\parallel.$ Neither of these necessarily has unit norm. Also let $\phi$ be the angle from $\unitvec{e}_1$ to $\myvector{c}$ and $\theta$ be the angle from $\myvector{c}$ to $\myvector{r}_\textrm{0}^\parallel.$ As these three vectors lie in the same plane the angles can be chosen so that the  from $\unitvec{e}_1$ to $\myvector{r}_\textrm{0}^\parallel$ is $\theta+\phi.$ Then vector algebra gives
\begin{subequations}
\begin{eqnarray}
	P_1 \projc \rinitial	& = & r_\textrm{0}^\parallel
	                                      \cos{\theta}\cos{\phi}\;
	                                      \unitvec{e}_1
																				\unitvec{e}_1 \quad \textrm{and}
																				\\ 
																				P_1(I - \projc)\rinitial & = & - r_\textrm{0}^\parallel
	                                      \sin{\theta}\sin{\phi}\;
\end{eqnarray}
\end{subequations}
where $r_\textrm{0}^\parallel$ is the magnitude of $\myvector{r}_\textrm{0}^\parallel.$ 

Then
\begin{subequations}
\begin{eqnarray}
 \rinitial^\top \projc  P_1 \projc\rinitial 
          &= & 
               \left( r_\textrm{0}^\parallel\right)^2
	             \cos^2{\theta}\cos^2{\phi}
							 \quad
							 \textrm{and}
							 \nonumber \\
   \rinitial^\top 
   \left( I - \projc \right) 
	 P_1
   \left( I - \projc \right)
	 \rinitial & = &\left( r_\textrm{0}^\parallel\right)^2
	             \sin^2{\theta}\sin^2{\phi}
\end{eqnarray}
\end{subequations}
Separately $\myvector{c}^\top P_1\myvector{c} = \cos^2{\phi}$  Thus 
\begin{eqnarray}
 W & = & \left( r_\textrm{0}^\parallel\right)^2 \left[ \sin^2{\theta}\sin^2{\phi} +(2-n) \cos^2{\theta}\cos^2{\phi} \right] 
          \nonumber \\
		 &  & + (n-1)\cos^2{\phi}
          \nonumber \\
		 & = & \left( r_\textrm{0}^\parallel\right)^2 \left\{ \cos^2{\theta}\left[(3-n)\cos^2{\phi} -1\right] - \cos^2{\phi} \right\} 
          \nonumber \\
		 &  & + (n-1)\cos^2{\phi}
\end{eqnarray}
For a given channel and choice of control direction $\myvector{c}$, the variable $\phi$ is fixed and this must be optimized with respect to $\rinitial$, i.e.\ with respect to $\theta$ and $r_\textrm{0}^\parallel$. Here, noting that for $n\geqslant 2$, the term $(3-n)\cos^2{\phi} -1 \leqslant 0.$ This implies that the factor multiplying $\left( r_\textrm{0}^\parallel\right)^2$ is never positive. So the maximum for $W_1$ is attained when $r_\textrm{0}^\parallel = 0.$ This gives $W_1 = (n-1)\cos^2{\phi}$ and this attains a maximum of $n-1$ when $\myvector{c}$ is perpendicular to $\unitvec{e}_1.$


\section{Higher order QFI terms for unital channels}
\label{app:higherorderqfiterms}

Equation~\eqref{eq:horders} and that facts that $\rhoforder{0} = \iop^{\otimes n}/N$ and $\scoreorder{0}=0$ imply that, to determine the third and fourth order terms in the QFI, we will need both $\rhoforder{j}$ and $\scoreorder{j}$ for $j=1,2,3.$ Then Eqs.~\eqref{eq:sldorders} and~\eqref{eq:unitalsld} and $\rhoforder{0} = \iop^{\otimes n}/N$ give 
\begin{equation}
	\frac{\partial \rhoforder{2}}{\partial \lambda}  = \frac{1}{N}\, \scoreorder{2}
	                                                      + \frac{N}{2} \left\{
																												                  \frac{\partial \rhoforder{1}}{\partial \lambda},\rhoforder{1} 
																																		  \right\} 
\end{equation}
where $\{ , \}$ indicates the anti-commutator. Thus
\begin{equation}
	\scoreorder{2}  = N \frac{\partial \rhoforder{2}}{\partial \lambda} 
	                                                      - \frac{N^2}{2}
																												  \frac{\partial \phantom{\lambda}}{\partial \lambda} \left[ \rhoforder{1} \right]^2. 
\end{equation}
Repeating this process gives
\begin{eqnarray}
	\frac{\partial \rhoforder{3}}{\partial \lambda}  & = & \frac{1}{N}\, \scoreorder{3}
																												\nonumber \\
	                                                 &   &      + \frac{N}{2} 
																									              \frac{\partial \phantom{\lambda}}{\partial \lambda} 
																																     \left\{
																												                  \rhoforder{2},\rhoforder{1} 
																																		  \right\}
																												\nonumber \\
																									 &  &  -\frac{N^2}{4}\left\{
																												                  \frac{\partial \phantom{\lambda}}{\partial \lambda} \left[ \rhoforder{1} \right]^2,\rhoforder{1} 
																																		  \right\}. 
\end{eqnarray}
Thus
\begin{eqnarray}
	\scoreorder{3}  & = & N \frac{\partial \rhoforder{3}}{\partial \lambda} 
	                                                      - \frac{N^2}{2}
																									              \frac{\partial \phantom{\lambda}}{\partial \lambda} 
																																     \left\{
																												                  \rhoforder{2},\rhoforder{1} 
																																		  \right\}
																												\nonumber \\
																									 &  &  +\frac{N^3}{4}\left\{
																												                  \frac{\partial \phantom{\lambda}}{\partial \lambda} \left[ \rhoforder{1} \right]^2,\rhoforder{1} 
																																		  \right\}. 
\end{eqnarray}

Then Eqs.~\eqref{eq:horders} and~\eqref{eq:unitalsld}, the fact that $\rhoforder{0} = \iop^{\otimes n}/N$ and the preceding expressions for the SLD terms give
\begin{eqnarray}
 \Horder{3} & = & \Trace{
                         \left[
												     \scoreorder{1} \frac{\partial \rhoforder{2}}{\partial \lambda} 
														 +
												     \scoreorder{2} \frac{\partial \rhoforder{1}}{\partial \lambda} 
                         \right]
												}
									\nonumber \\
						& = & 2N \Trace{
                         \left[
												     \frac{\partial \rhoforder{1}}{\partial \lambda} 
												     \frac{\partial \rhoforder{2}}{\partial \lambda} 
                         \right]
												} 
									\nonumber \\
						&  &
										-\frac{N^2}{2}\Trace{
                         \left[
												       \frac{\partial \left(\rhoforder{1} \right)^2}{\partial \lambda} 
															 \frac{\partial \rhoforder{1}}{\partial \lambda} 
                         \right]
												}.
\end{eqnarray} 

Thus
\begin{eqnarray}
 \Horder{3} & = & 
                 2N \Trace{
                         \left[
												     \frac{\partial \rhoforder{1}}{\partial \lambda} 
												     \frac{\partial \rhoforder{2}}{\partial \lambda} 
                         \right]
												} 
									\nonumber \\
						&  &
										-N^2\Trace{
                         \left[
												       \frac{\partial \left(\rhoforder{1} \right)^2}{\partial \lambda} 
															 \rhoforder{1} 
                         \right].
												}
\label{eq:horderthree}
\end{eqnarray}

The fourth order term in the QFI can be obtained in a similar fashion, eventually giving
\begin{eqnarray}
 \Horder{4} & = & 
                 N \Trace{
                         \left[
												     2
														 \frac{\partial \rhoforder{1}}{\partial \lambda} 
												     \frac{\partial \rhoforder{3}}{\partial \lambda} 
														 + 
														 \left(
														   \frac{\partial \rhoforder{2}}{\partial \lambda}  
														\right)^2
                         \right]
												} 
									\nonumber \\
						&  &
										-\frac{N^2}{2}\, 
										  \Trace{
                         \left[
												       \frac{\partial \rhoforder{1}}{\partial \lambda}
															 \frac{\partial \phantom{\lambda}}{\partial \lambda}\
															 \left\{
															  \rhoforder{1}, \rhoforder{2} 
															 \right\} 
															+
												       \frac{\partial \rhoforder{2}}{\partial \lambda}
															 \frac{\partial \left(
															  \rhoforder{1} 
															 \right)^2 }{\partial \lambda}
                         \right]
												}
									\nonumber \\
						&  & + \frac{N^3}{4}\,
						       \Trace{
									  \left[ 
										  \frac{\partial \rhoforder{1}}{\partial \lambda}
										\left\{
											\frac{\partial \phantom{\lambda}}{\partial \lambda} \left[ \rhoforder{1} \right]^2,\rhoforder{1} 
										\right\}
										\right]
										}.
\end{eqnarray}
Further algebra yields
\begin{eqnarray}
 \Horder{4} & = & 
                 N \Trace{
                         \left[
												     2
														 \frac{\partial \rhoforder{1}}{\partial \lambda} 
												     \frac{\partial \rhoforder{3}}{\partial \lambda} 
														 + 
														 \left(
														   \frac{\partial \rhoforder{2}}{\partial \lambda}  
														\right)^2
                         \right]
												} 
									\nonumber \\
						&  &
										-N^2\, 
										  \Trace{
                         \left[
												       \left( \frac{\partial \rhoforder{1}}{\partial \lambda}\right)^2
															 \rhoforder{2} 
															+
												       \frac{\partial \rhoforder{2}}{\partial \lambda}
															 \frac{\partial \left(
															  \rhoforder{1} 
															 \right)^2 }{\partial \lambda}
                         \right]
												}
									\nonumber \\
						&  & + \frac{N^3}{4}\,
						       \Trace{
									  \left[ 
										 \left( \frac{\partial \left(
															  \rhoforder{1} 
															 \right)^2 }{\partial \lambda}\right)^2
										\right]
										}.
\label{eq:horderfour}
\end{eqnarray}

The second and third order terms in the channel input state can be determined by repeatedly using Eq.~\eqref{eq:ucactionone} and the additional result
\begin{eqnarray}
  \uc \left( \sigmagen{a} \otimes \sigmagen{b} \right) \ucconj & = & \left( \myvector{a} \times \myvector{c} \right) \cdot \hat{\myvector{\sigma}}
                                                                    \otimes 
																																		\left( \myvector{b} \times \myvector{c} \right) \cdot \hat{\myvector{\sigma}}
																																		\nonumber \\
                                                      &   & + \left( \myvector{a}\cdot\myvector{c} \right)\, \iop \otimes \sigmagen{b}
																														+ \left( \myvector{b}\cdot\myvector{c} \right)\, \sigmagen{a} \otimes \iop 
																																		\nonumber \\
                                                      &   & + \left( \myvector{a}\cdot\myvector{c} \right) 
																											        \left( \myvector{b}\cdot\myvector{c} \right)
																						                  \bigl( \sigmagen{c} \otimes \sigmagen{c} \bigr. 
																																		\nonumber \\
                                                      &   &  
																															 \bigl. - \sigmagen{c} \otimes \iop  - \iop \otimes \sigmagen{c} \bigr)
                                                             \label{eq:ucactiontwo}. 
\end{eqnarray}
To prove this, note that 
   $\uc \left( \sigmagen{a} \otimes \sigmagen{b} \right) \ucconj = 
     \uc \left( \sigmagen{a} \otimes \iop \right) 
		 \uc \ucconj
		\left( \iop  \otimes \sigmagen{b}\right)\ucconj$
since $\uc \ucconj = \iop.$
Then invoking~\eqref{eq:ucactionone} twice and using~\eqref{eq:paulimulti} repeatedly gives~\eqref{eq:ucactiontwo}.  

We apply these to determine the second and third order terms in the channel input density operator provided that \emph{the initial state Bloch sphere direction and the control direction are perpendicular.} Eq.~\eqref{eq:corrorderoneinput} implies that 
\begin{eqnarray}
\rhoiorder{1} & = &
                \frac{1}{N}\;
								\left(
								  \sigmagen{\rinitial} \otimes \sigmagen{c}^{\otimes(n-1)}
								\right.
									\nonumber \\
							&   & 
							  \left.
									+ \cdots + 
									\sigmagen{c}^{\otimes(n-1)} \otimes \sigmagen{\rinitial}
								\right)
\end{eqnarray}
and this contains all terms with a single factor of $\sigmagen{\rinitial}$ and $n-1$ factors of $\sigmagen{c}$.

The second order term in the initial state is
\begin{eqnarray}
  \rhooorder{2} & = & \frac{1}{N}\, \left[ \sigmagen{\rinitial} \otimes \sigmagen{\rinitial} \otimes \iop^{\otimes(n-2)}
	                    \right. 
											\nonumber \\
	              &   & \left.
																+ \ldots
																			 + \iop^{\otimes(n-2)} \otimes \sigmagen{\rinitial} \otimes \sigmagen{\rinitial}
	                              \right]
\end{eqnarray}
where every possible permutation including exactly two factors of $\sigmagen{\rinitial}$ appears with the brackets. Then consider the effect of the preparatory unitary on $\sigmagen{\rinitial} \otimes \sigmagen{\rinitial} \otimes \iop^{\otimes(n-2)}$. We label the qubits from left to right as $1,2,3,\ldots, n$. Then $\uc$ acting on any pair of qubits in the range $3,4,\ldots, n$ has no effect on this term. Now consider $\uc$ acting on qubits $1$ and $3$. Since $\rinitial$ is perpendicular to $\myvector{c}$, Eq.~\eqref{eq:ucactionone} shows that this produces $\sigmagen{\rinitial} \otimes \sigmagen{\rinitial} \otimes \sigmagen{c} \otimes \iop^{\otimes(n-3)}$. Now consider the subsequent action of $\uc$ on qubits $2$ and $3$. According to Eq.~\eqref{eq:ucactiontwo} this results in $\sigmagen{\rinitial} \otimes \sigmagen{\rinitial} \otimes \iop^{\otimes(n-2)}$. Thus the only factor of $\uc$ that will have a non-trivial effect on this terms is that acting on qubits $1$ and $2$. Eq.~\eqref{eq:ucactiontwo} shows that this gives $\sigmagen{\rinitial \times \myvector{c}} \otimes \sigmagen{\rinitial \times \myvector{c}} \otimes \iop^{\otimes(n-2)}$. Thus, if $\rinitial$ is perpendicular to $\myvector{c},$ the second order channel input term is 
\begin{eqnarray}
  \rhoiorder{2} & = & \frac{1}{N}\, \left[ \sigmagen{\rinitial \times \myvector{c}} \otimes \sigmagen{\rinitial \times \myvector{c}} \otimes \iop^{\otimes(n-2)}
	                    \right. 
											\nonumber \\
	              &   & \left.
										  +\ldots
											+\iop^{\otimes(n-2)} \otimes \sigmagen{\rinitial \times \myvector{c}} \otimes \sigmagen{\rinitial \times \myvector{c}} 
	                              \right]
\end{eqnarray}
which contains terms with every possible arrangement of two factors of $\sigmagen{\rinitial \times \myvector{c}}$.

The third order initial state term is
\begin{eqnarray}
  \rhooorder{3} & = & \frac{1}{N}\, \left[ \sigmagen{\rinitial} \otimes \sigmagen{\rinitial} \otimes \sigmagen{\rinitial} \otimes \iop^{\otimes(n-3)}
	                    \right. 
											\nonumber \\
	              &   & \left.
																+ \ldots
																			 + \iop^{\otimes(n-3)} \otimes \sigmagen{\rinitial} \otimes \sigmagen{\rinitial} \otimes \sigmagen{\rinitial}
	                              \right]
\end{eqnarray}
and this contains terms with every possible arrangement of three factors of $\sigmagen{\rinitial}$. The effects of the preparatory unitary on this can be determined by considering $\sigmagen{\rinitial} \otimes \sigmagen{\rinitial} \otimes \sigmagen{\rinitial} \otimes \iop^{\otimes(n-3)}$. Then by the argument for the second order term the factors of $\uc$ acting on all pairs of qubits from $2$ to $n$ produce $\sigmagen{\rinitial} \otimes \sigmagen{\rinitial \times \myvector{c}} \otimes \sigmagen{\rinitial \times \myvector{c}} \otimes \iop^{\otimes(n-3)}$. It remains to consider the effect of each $\uc$ which involves qubit~$1$. By Eq.~\eqref{eq:ucactionone} the effect of $\uc$ for qubits $1$ and $4$, $\uc$ for qubits $1$ and $5$ up to $\uc$ for qubits $1$ and $n$ is to produce $\sigmagen{\rinitial} \otimes \sigmagen{\rinitial \times \myvector{c}} \otimes \sigmagen{\rinitial \times \myvector{c}} \otimes \sigmagen{c}^{\otimes(n-3)}$. Now the effect of $\uc$ for qubits $1$ and $3$ on this is to produce $\sigmagen{\rinitial \times \myvector{c}} \otimes \sigmagen{\rinitial \times \myvector{c}} \otimes \sigmagen{(\rinitial \times \myvector{c})  \times \myvector{c} }  \otimes \sigmagen{c}^{\otimes(n-3)}$ which is the same as $-\sigmagen{\rinitial \times \myvector{c}} \otimes\sigmagen{\rinitial \times \myvector{c}}  \otimes \sigmagen{\rinitial}  \otimes \sigmagen{c}^{\otimes(n-3)}$. Finally the effect of $\uc$ for qubits $1$ and $2$ on this is to produce $-\sigmagen{\rinitial} \otimes\sigmagen{\rinitial}  \otimes \sigmagen{\rinitial}  \otimes \sigmagen{c}^{\otimes(n-3)}$. Thus we obtain the third order term in the channel input state, 
\begin{eqnarray}
  \rhoiorder{3} & = & -\frac{1}{N}\, \left[ \sigmagen{\rinitial} \otimes \sigmagen{\rinitial} \otimes \sigmagen{\rinitial} \otimes \sigmagen{c}^{\otimes(n-3)}
	                    \right. 
											\nonumber \\
	              &   & \left.
																+ \ldots
																			 + \sigmagen{c}^{\otimes(n-3)} \otimes \sigmagen{\rinitial} \otimes \sigmagen{\rinitial} \otimes \sigmagen{\rinitial}
	                              \right]
\end{eqnarray}
where this contains all terms with three factors of $\sigmagen{\rinitial}$.

We now consider the situation where the channel acts on once on the leftmost qubit and aim to compute the third and fourth orders terms in the QFI via Eqs.~\eqref{eq:horderthree} and~\eqref{eq:horderfour}. The key tools are that $\Trace{\left[\sigmagen{a}\right]}=0$, $\Trace{\left[\sigmagen{a} \sigmagen{b}\right]} = 2\myvector{a} \cdot \myvector{b}$ and that the trace of a product state is the product of the partial traces of the factors. These immediately imply that all terms in $\Horder{3}$ and $\Horder{4}$ that contain a factor of $\rhoforder{2}$ or $\frac{\partial \rhoforder{2}}{\partial \lambda}$ multiplied by factors that contain $\rhoforder{1}$ or its derivative trace to zero, since these always contain factors of $\Trace{\left[ \sigmagen{\rinitial \times c} \right]}$, $\Trace{\left[ \sigmagen{\rinitial \times c} \sigmagen{\rinitial} \right]}$ or $\Trace{\left[ \sigmagen{\rinitial \times c} \sigmagen{c} \right]}.$
Similarly the product $\frac{\partial \rhoforder{1}}{\partial \lambda} \frac{\partial \rhoforder{3}}{\partial \lambda}$ contains at least one factor of $\sigmagen{\rinitial}$ and this also traces to zero. Thus, in this scenario,
\begin{subequations}
\begin{eqnarray}
  \Horder{3}  & = &
										-N^2\Trace{
                         \left[
												       \frac{\partial \left(\rhoforder{1} \right)^2}{\partial \lambda} 
															 \rhoforder{1} 
                         \right]
												}
									  \; \textrm{and} 
                    \label{eq:horderthreesimp} \\
 \Horder{4}   & = & 
                 N \Trace{
                         \left[
														 \left(
														   \frac{\partial \rhoforder{2}}{\partial \lambda}  
														\right)^2
                         \right]
												} 
									\nonumber \\
						&  & + \frac{N^3}{4}\,
						       \Trace{
									  \left[ 
										 \left( \frac{\partial \left(
															  \rhoforder{1} 
															 \right)^2 }{\partial \lambda}\right)^2
										\right]
										}.
   \label{eq:horderfoursimp}
\end{eqnarray}
\end{subequations}

Now consider $\frac{\partial \left(\rhoforder{1} \right)^2}{\partial \lambda} \rhoforder{1}$. Within the rightmost $n-1$ factors there will always be at least one factor of $\sigmagen{c}$ or $\sigmagen{\rinitial}$. Thus this traces to zero and Eq.~\eqref{eq:horderthreesimp} gives that if $\rinitial$ and $\myvector{c}$ are perpendicular then $\Horder{3} =0.$

In order to compute  $\Horder{4}$, we first consider 
$ \left(
		\frac{\partial \rhoforder{2}}{\partial \lambda}  
\right)^2$. Here 
\begin{eqnarray}
  \frac{\partial \rhoforder{2}}{\partial \lambda} & = & \frac{1}{N}\, 
	                    \left[ \dot{M} (\rinitial \times \myvector{c}) \cdot \hat{\myvector{\sigma}} 
											       \otimes 
														 \sigmagen{\rinitial \times c} \otimes \iop^{\otimes(n-2)}
											\right. 
											\nonumber \\
							 &  & + \dot{M} (\rinitial \times \myvector{c}) \cdot \hat{\myvector{\sigma}} 
											       \otimes 
														 \iop 
											       \otimes 
														 \sigmagen{\rinitial \times c} \otimes \iop^{\otimes(n-3)}
											\nonumber \\
							 &  & + \left.
							         \ldots + \dot{M} (\rinitial \times \myvector{c}) \cdot \hat{\myvector{\sigma}} 
											       \otimes 
														 \iop^{\otimes(n-2)} 
											       \otimes 
														 \sigmagen{\rinitial \times c}
											\nonumber 
										\right].
\end{eqnarray}
After squaring, the cross terms from this each contain at least one factor of $\sigmagen{\rinitial \times c}$ and do not contribute to the trace. The $n-1$ squared terms are all $(\rinitial \times \myvector{c})^\top \dot{M}^\top \dot{M}(\rinitial \times \myvector{c})  \iop^{\otimes n}/N^2$. Thus
\begin{equation}
 \Trace{
                         \left[
														 \left(
														   \frac{\partial \rhoforder{2}}{\partial \lambda}  
														\right)^2
                         \right]
												} 
	= \frac{n-1}{N}\,
	  (\rinitial \times \myvector{c})^\top \dot{M}^\top \dot{M}(\rinitial \times \myvector{c}).
\end{equation}

The remaining term involves
\begin{eqnarray}
  \rhoforder{1} & = & \frac{1}{N}\, 
	                    \left[ \left( M \rinitial \right) \cdot \hat{\myvector{\sigma}} 
											       \otimes 
														 \sigmagen{c}^{\otimes(n-1)}
											\right. 
											\nonumber \\
                & & + \left( M \myvector{c} \right) \cdot \hat{\myvector{\sigma}} 
											       \otimes 
														 \sigmagen{\rinitial}
											       \otimes 
														 \sigmagen{c}^{\otimes(n-2)}
											\nonumber \\
                & & \left. + \ldots + 
								    \left( M \myvector{c} \right) \cdot \hat{\myvector{\sigma}} 
											       \otimes 
														 \sigmagen{c}^{\otimes(n-2)}
														 \otimes
														 \sigmagen{\rinitial}
										\right].
\end{eqnarray}
Thus for $n \geqslant 3$,
\begin{eqnarray}
  \left(\rhoforder{1}\right)^2
	             & = & \frac{1}{N^2}\, 
	                    \left[ \rinitial^\top M^\top M \rinitial\, 
														 \iop^{\otimes n}
											+ (n-1) \myvector{c}^\top M^\top M \myvector{c} \, 
														 \iop^{\otimes n}
											\right. 
											\nonumber \\
	             &  & + (n-1) \myvector{c}^\top M^\top M \myvector{c} \,
							              \left(
														 \iop
														 \otimes 
														 \sigmagen{\rinitial \times c} 
														 \otimes 
														 \sigmagen{\rinitial \times c} 
														 \otimes
														 \iop^{\otimes (n-3)}
														\right.
											\nonumber \\
								& & + \ldots
								            \left. 
														 \iop^{\otimes(n-2)} 
														 \otimes
														 \sigmagen{\rinitial \times c} 
														 \otimes
														 \sigmagen{\rinitial \times c}
														\right)
											\nonumber \\
                & & + \left( M \myvector{\rinitial} \times M \myvector{c} \right) \cdot \hat{\myvector{\sigma}} 
											\otimes 
														 \sigmagen{\rinitial \times c}
											       \otimes 
														 \iop^{\otimes(n-2)}
											\nonumber \\
                & & \left.
								    + \ldots + \left( M \myvector{\rinitial} \times M \myvector{c} \right) \cdot \hat{\myvector{\sigma}} 
											\otimes
											\iop^{\otimes(n-2)}
											\otimes
											\sigmagen{\rinitial \times c}
										\right].
\end{eqnarray}
After differentiating and squaring, only the squares contribute to the trace. Thus
\begin{eqnarray}
  \Trace{\left[ 
				 \left( \frac{\partial \left(
				  \rhoforder{1} 
         \right)^2 }{\partial \lambda}\right)^2
				\right]
        }
				& = &
				\frac{1}{N^3}\,
				\left[ \rinitial^\top \frac{\partial \left( M^\top M\right) }{\partial \lambda} \rinitial
				\right.
				\nonumber \\
				& & 
				\left.
				       +(n-1)\, \myvector{c}^\top \frac{\partial \left( M^\top M\right) }{\partial \lambda} \myvector{c}
				\right]^2
				\nonumber \\
				& & + \frac{4(n-1)}{N^3} 
				      \left\lvert 
							   \frac{\partial}{\partial \lambda}
								 \left(
								   M\rinitial \times M \myvector{c}
								 \right)
							\right\rvert^2
				\nonumber \\
				& & + \frac{2(n-1)(n-2)}{N^3} 
				      \left[ 
							   \myvector{c}^\top \frac{\partial \left( M^\top M\right) }{\partial \lambda} \myvector{c}
							\right]^2
				\nonumber \\
				& &
\end{eqnarray}
where $\lvert \myvector{a} \rvert^2 := \myvector{a} \cdot \myvector{a}.$ Thus
\begin{eqnarray}
  \Horder{4} & = & (n-1)\, 
	                 (\rinitial \times \myvector{c})^\top \dot{M}^\top \dot{M}(\rinitial \times \myvector{c})
									\nonumber \\
						 &   & 
				\frac{1}{4}\,
				\left[ \rinitial^\top \frac{\partial \left( M^\top M\right) }{\partial \lambda} \rinitial
				\right.
				\nonumber \\
				& & 
				\left.
				       +(n-1)\, \myvector{c}^\top \frac{\partial \left( M^\top M\right) }{\partial \lambda} \myvector{c}
				\right]^2
				\nonumber \\
				& & + (n-1)\, 
				      \left\lvert 
							   \frac{\partial}{\partial \lambda}
								 \left(
								   M\rinitial \times M \myvector{c}
								 \right)
							\right\rvert^2
				\nonumber \\
				& & + \frac{(n-1)(n-2)}{2} 
				      \left[ 
							   \myvector{c}^\top \frac{\partial \left( M^\top M\right) }{\partial \lambda} \myvector{c}
							\right]^2.
				\nonumber \\
				& &
\end{eqnarray}
This scales as $n^2$. For example, for the depolarizing channel, $M=\lambda I$ and thus
\begin{equation}
  \Horder{4} = (n-1) + \lambda^2 n (3n-2).
\end{equation}


\section{Measurements that saturate the QFI}
\label{app:measurements}

Suppose that the single channel invocation is followed by application of $\uprep^\dagger = \uprep$ and, after this each qubit is measured in a particular basis. We will consider the situation where $\rinitial$ and $\myvector{c}$ are perpendicular and show that for a particular set of single qubit measurements, the resulting classical Fisher information gives the bounds of Eq.~\eqref{eq:corrqfibounds}. For $\rinitial$ and $\myvector{c}$ perpendicular, prior to channel invocation the state of the system is, to first order in the purity,
\begin{equation}
 \rhoi = \rhoiorder{0} + r \rhoiorder{1}
\end{equation}
where $\rhoiorder{0} = \iop ^{\otimes n}/N$ and 
\begin{equation}
 \rhoiorder{1} = 
                \frac{1}{N}\;
								\left(
								  \sigmagen{\rinitial} \otimes \sigmagen{c}^{\otimes(n-1)}
									+ \cdots + 
									\sigmagen{c}^{\otimes(n-1)} \otimes \sigmagen{\rinitial}
								\right) 
\end{equation}
Here the sum contains every possible term with only one factor of $\sigmagen{\rinitial}$.

Under a unital channel, $\iop$ remains invariant and thus a single channel invocation on the leftmost qubit results in the state
\begin{equation}
 \rhof = \rhoforder{0} + r \rhoforder{1}
\end{equation}
where $\rhoforder{0} = \iop ^{\otimes n}/N$ and 
\begin{eqnarray}
 \rhoforder{1} & = &
                \frac{1}{N}\;
								\left(
								  \sigmagen{a} \otimes \sigmagen{c}^{\otimes(n-1)}
									+ 
								\right.
								\nonumber \\
							& & +
							  \sigmagen{b} \otimes \sigmagen{\rinitial} \otimes \sigmagen{c}^{\otimes(n-2)}
								\nonumber \\
							& & \left.
							\cdots + 
									\sigmagen{b} \otimes \sigmagen{c}^{\otimes(n-2)} \otimes \sigmagen{\rinitial}
								\right). 
\end{eqnarray}
Here, for convenience, $\myvector{a} := M \rinitial$ and $\myvector{b} := M \myvector{c}.$ 

The pre-measurement state of the system is then $\rhom:= \uprep^\dagger \rhof \uprep$ and 
\begin{equation}
 \rhom = \rhomorder{0} + r \rhomorder{1}
\end{equation}
where $\rhomorder{i} := \uprep^\dagger \rhoforder{i} \uprep$. Then 
\begin{equation}
 \rhomorder{0} = \frac{1}{N}\; \iop ^{\otimes n}
\label{eq:rhomeasureorderzero}
\end{equation}
since $\uprep$ is unitary. 

In order to determine $\rhomorder{1}$, note that $\uprep$ is a product of one $\uc$ for each pair of qubits. We will determine $\rhomorder{1}$ in two steps: a) evaluate the effect of every factor of $\uc$ that does not include the leftmost qubit and b) evaluate the effect of every factor of $\uc$ that does contain the leftmost qubit. 

First, consider the effect of every factor of $\uc$ that does not include the leftmost qubit. We need to consider two types of terms in $\rhoforder{1}$. One type is $\sigmagen{a} \otimes \sigmagen{c}^{\otimes(n-1)}$ and this stays invariant. The other type contains a single factor of $\sigmagen{\rinitial}$ in the rightmost $n-1$ factors of the tensor product. As an example consider $\sigmagen{b} \otimes \sigmagen{\rinitial} \otimes \sigmagen{c}^{\otimes(n-2)}$. Determining the effect of the factors of $\uc$ requires only determining the effect of the factors that involve the second qubit from the left. Then using Eq.~\eqref{eq:ucactiontwo} and the fact that $\rinitial$ and $\myvector{c}$ are perpendicular gives
\begin{equation}
  \uc \sigmagen{\rinitial} \otimes \sigmagen{c} \uc^\dagger = \sigmagen{\rinitial} \otimes \iop.
\end{equation}
Repeatedly using this gives that every factor of $\uc$ that does not include the leftmost qubit maps 
\begin{equation}
 \sigmagen{b} \otimes \sigmagen{\rinitial} \otimes \sigmagen{c}^{\otimes(n-2)} 
 \mapsto 
 \sigmagen{b} \otimes \sigmagen{\rinitial} \otimes \iop^{\otimes(n-2)}. 
\end{equation}
Thus every factor of $\uc$ that does not include the leftmost qubit maps
\begin{eqnarray}
 \rhoforder{1} & \mapsto &
                \frac{1}{N}\;
								\left(
								  \sigmagen{a} \otimes \sigmagen{c}^{\otimes(n-1)}
									+ 
								\right.
								\nonumber \\
							& & +
							  \sigmagen{b} \otimes \sigmagen{\rinitial} \otimes \iop^{\otimes(n-2)}
								\nonumber \\
							& & \left.
							\cdots + 
									\sigmagen{b} \otimes \iop^{\otimes(n-2)} \otimes \sigmagen{\rinitial}
								\right). 
\label{eq:rhomeasureoneint}
\end{eqnarray}

Second, consider the remaining factors of $\uc$ that each include the leftmost qubit. In order to assess these, we use Eq.~\eqref{eq:ucactiontwo} to give
\begin{eqnarray}
 \uc \sigmagen{a} \otimes \sigmagen{c} \uc^\dagger & = & \sigmagen{a} \otimes \iop
																																		\nonumber \\
                                                         & &
                                                         + \left( \myvector{a}\cdot \myvector{c} \right)
																												   \left( \sigmagen{c} \otimes \sigmagen{c} - \sigmagen{c} \otimes  \iop \right),
																												 \\
 \uc \sigmagen{b} \otimes \sigmagen{\rinitial} \uc^\dagger & = & \left( \myvector{b} \times \myvector{c} \right) \cdot \hat{\myvector{\sigma}}
                                                                    \otimes 
																																		\left( \myvector{\rinitial} \times \myvector{c} \right) \cdot \hat{\myvector{\sigma}}
																																		\nonumber \\
                                                         & & + \left( \myvector{b}\cdot \myvector{c} \right) \,  \iop \otimes \sigmagen{\rinitial } \quad \textrm{and}
																												 \\
 \uc \sigmagen{b} \otimes \iop \uc^\dagger & = & \sigmagen{b} \otimes  \sigmagen{c}
																																		\nonumber \\
                                                         & & 
                                                         + \left( \myvector{b}\cdot \myvector{c} \right)
																												   \left( \sigmagen{c} \otimes  \iop  - \sigmagen{c} \otimes  \sigmagen{c} \right).
\end{eqnarray}
Repeatedly applying these gives that under $\uc$ acting on every pair that includes the leftmost qubit, 
\begin{eqnarray}
 \sigmagen{a} \otimes \sigmagen{c}^{\otimes(n-1)} & \mapsto & \sigmagen{a} \otimes \iop^{\otimes(n-1)}
                                                              + \left( \myvector{a}\cdot \myvector{c} \right) 
																													     \sigmagen{c}^{\otimes n}
                                                              \nonumber \\
																									        & & - \left( \myvector{a}\cdot \myvector{c} \right)  \sigmagen{c}\otimes \iop ^{\otimes (n-1)}.
\end{eqnarray}
The remaining terms can be evaluated by considering $\sigmagen{b} \otimes \sigmagen{\rinitial} \otimes \iop^{\otimes(n-2)}$. Again applying the results of Eq.~\eqref{eq:ucactiontwo} gives
\begin{eqnarray}
 \sigmagen{b} \otimes \sigmagen{\rinitial} \otimes \iop^{\otimes(n-2)} & \mapsto &
                                                                       \left( \myvector{b} \times \myvector{c} \right) \cdot \hat{\myvector{\sigma}}
                                                                       \otimes 
																																		   \left( \myvector{\rinitial} \times \myvector{c} \right) \cdot \hat{\myvector{\sigma}}
																																			 \otimes \sigmagen{c}^{\otimes (n-2)}
																																			 \nonumber \\
																																			& &  \left( \myvector{b}\cdot \myvector{c} \right)
																																			     \iop \otimes \sigmagen{\rinitial} \otimes \iop^{\otimes (n-2)}.
\end{eqnarray}

Extending these results to all the permutations in Eq.~\eqref{eq:rhomeasureoneint} then gives
\begin{eqnarray}
 \rhomorder{1} & = & \frac{1}{N} \left[ \sigmagen{a} \otimes \iop^{\otimes(n-1)} 
                                 \right.
																 \nonumber \\
							 & & \left( \myvector{a}\cdot \myvector{c} \right) 
									 \left( \sigmagen{c}^{\otimes n} - \sigmagen{c}\otimes \iop ^{\otimes (n-1)} \right)
                   \nonumber \\
							& &  \left( \myvector{b} \times \myvector{c} \right) \cdot \hat{\myvector{\sigma}}
									 \otimes 
									 \left( \myvector{\rinitial} \times \myvector{c} \right) \cdot \hat{\myvector{\sigma}}
									 \otimes \sigmagen{c}^{\otimes (n-2)}
                   \nonumber \\
							& &  + \cdots + \left( \myvector{b} \times \myvector{c} \right) \cdot \hat{\myvector{\sigma}}
									 \otimes \sigmagen{c}^{\otimes (n-2)}
									 \otimes 
									 \left( \myvector{\rinitial} \times \myvector{c} \right) \cdot \hat{\myvector{\sigma}}
									 \nonumber \\
							& &  \left( \myvector{b} \cdot \myvector{c} \right)
							     \left( \iop \otimes \sigmagen{\rinitial} \otimes \iop^{\otimes(n-2)} \right.
									 \nonumber \\
							& &  \left. \left.
									        + \cdots + \iop^{\otimes(n-1)} \otimes \sigmagen{\rinitial}
									 \right) \right].
\label{eq:rhomeasureorderone}
\end{eqnarray}

Now consider measurement for each qubit along the Bloch sphere direction $\rinitial$. The corresponding single qubit projection operators are
\begin{equation}
 \hat{\Pi}_\pm := \frac{1}{2}\, \left( \iop \pm \sigmagen{\rinitial} \right)
\end{equation}
and we denote the associated measurement outcomes by $+$ or $-$. Since $\rhomorder{1}$ is invariant under interchange between any of the rightmost $n-1$, the probabilities of the measurement outcomes depend on: a) whether the outcome for the leftmost qubit is $+$ or $-$ and b) the number, $k$, of $+$ outcomes amongst the rightmost $n-1$ qubits. To this end let $p(+,k)$ be the probability that the measurement outcome for the leftmost qubit is $+$ and $k$ of the outcomes amongst the rightmost $n-1$ qubits are $+$. Similarly let $p(-,k)$ be the probability that the measurement outcome for the leftmost qubit is $-$ and $k$ of the outcomes amongst the rightmost $n-1$ qubits are $+$.

A simplifying aspect of computing these probabilities is the fact that if $\rinitial$ and $\myvector{c}$ are perpendicular then $\Trace{\left( \hat{\Pi}_\pm \sigmagen{c} \right)} =0.$ Thus only the first and last series of terms on the right hand side of Eq.~\eqref{eq:rhomeasureorderone} contribute to the probabilities. Now consider $p(+,k)$. One way of attaining this outcome is a $+$ for the leftmost qubit, followed by $+$ outcomes for the next $k$ qubits and finally $-$ outcomes for the rightmost $n-1-k$ qubits. The probability with which this occurs is
\begin{eqnarray}
 & & \frac{1}{2^n}\, \Trace{\left[ \left( \iop + \sigmagen{\rinitial} \right) 
                               \otimes \cdots \otimes
															 \left( \iop + \sigmagen{\rinitial} \right)
														 \right.} 
								               \nonumber \\
 & &                         \left.
                                 \otimes
                                 \left( \iop - \sigmagen{\rinitial} \right)
                                 \otimes \cdots \otimes 
																 \left( \iop - \sigmagen{\rinitial} \right)
																 \rhom \right]. 
\end{eqnarray}
where there are $k+1$ factors with the $+$ sign and $n-k-1$ with the $-$ sign. Using Eqs.~\eqref{eq:rhomeasureorderzero} and ~\eqref{eq:rhomeasureorderone} then gives that this returns 
\begin{equation}
  \frac{1}{N}\,
	\left[ 
	 1 + r \left( \rinitial \cdot \myvector{a} \right) 
	  + r \left( \rinitial \cdot \myvector{c} \right) \left( 2k-n+1 \right)   
	\right].
\end{equation}
This, the fact that there are $\binom{n-1}{k}$ ways to attain $k$ outcomes of $+$ amongst the rightmost $n-1$ qubits and the definitions of $\myvector{a}$ and $\myvector{b}$ give
\begin{eqnarray}
 p(+,k) & = & \frac{1}{N} \binom{n-1}{k} 
              \left[
							 1 + r \rinitial^\top M \rinitial
              \right. 
							\nonumber \\
				& &  \left.
				      + r \myvector{c}^\top M \myvector{c} \left( 2k-n+1 \right)
						 \right].
\end{eqnarray} 
Similarly
\begin{eqnarray}
 p(-,k) & = & \frac{1}{N} \binom{n-1}{k} 
              \left[
							 1 - r \rinitial^\top M \rinitial
              \right. 
							\nonumber \\
				& &  \left.
				      + r \myvector{c}^\top M \myvector{c} \left( 2k-n+1 \right)
						 \right].
\end{eqnarray} 

The classical Fisher information in this case is
\begin{eqnarray}
 F(\lambda) &= & \sum_{k=0}^{n-1} 
     \left[
		  \frac{1}{p(+,k)}\,
		  \left( \frac{\partial p(+,k)}{\partial \lambda}\right)^2
			\right.
			\nonumber \\
	& & 
			\left.
			+
		  \frac{1}{p(-,k)}\,
		  \left( \frac{\partial p(-,k)}{\partial \lambda}\right)^2
     \right].
\end{eqnarray}
Substituting and retaining only the lowest order non-zero terms in the purity gives
\begin{equation}
 F(\lambda) = r^2 \left[ \left( \rinitial^\top \dot{M} \rinitial \right)^2
           + 
           (n-1) \left( \myvector{c}^\top \dot{M} \myvector{c} \right)^2
         \right].
\end{equation} 

Now consider the choices for $\rinitial$ and $\myvector{c}$ that resulted in the bounds of Eq.~\eqref{eq:corrqfibounds}. Following the notation of appendix~\ref{app:qfibound}, choosing $\myvector{c} = B^\top \unitvec{e}_1$ and $\rinitial = B^\top \unitvec{e}_2$, gives
\begin{equation}
 F(\lambda) = r^2 (n-1) s_1^2 + r^2 s_2^2. 
\end{equation} 
Thus this particular set of Bloch sphere direction and measurement choices saturates the lower bound of the QFI.


%

\end{document}